\documentclass[fleqn,usenatbib]{mnras}
\pdfoutput=1

\usepackage[T1]{fontenc}

\usepackage{hyperref}
\DeclareRobustCommand{\VAN}[3]{#2}
\let\VANthebibliography\thebibliography
\def\thebibliography{\DeclareRobustCommand{\VAN}[3]{##3}\VANthebibliography}

\usepackage{graphicx}	
\usepackage{amsmath}	
\usepackage{amssymb}	
\usepackage{newtxtext,newtxmath}

\title[Reflection spectra from accretion disc]{Emission lines from X-ray illuminated  accretion disc in black hole binaries}

\author[Mondal et al.]{
Santanu Mondal,$^{1,2}$\thanks{E-mail: santanuicsp@gmail.com}
Tek P. Adhikari,$^{3}$\thanks{E-mail: tek@iucaa.in}
Chandra B. Singh$^{4}$\thanks{E-mail: chandrasingh@ynu.edu.cn}
\\
$^{1}$Indian Institute of Astrophysics, II Block, Koramangala, Bengaluru 560034, India \\
$^{2}$Physics Department, Ben-Gurion University of the Negev, Be'er-Sheva 84105, Israel\\
$^{3}$Inter-University Center for Astronomy and Astrophysics (IUCAA), Pune 411007, India\\
$^{4}$South-Western Institute for Astronomy Research, Yunnan University, University Town, Chenggong District, Kunming 650500, P.R. China\\
}

\date{Accepted XXX. Received YYY; in original form ZZZ}

\pubyear{2020}

\begin{document}
\label{firstpage}
\pagerange{\pageref{firstpage}--\pageref{lastpage}}
\maketitle

\begin{abstract}
X-ray flux from the inner hot region around central compact object in a binary system illuminates the upper surface of an accretion disc and it behaves like a corona. This region can be photoionised by the illuminating radiation, thus can emit different emission lines. We study those line spectra in black hole X-ray binaries for different accretion flow parameters including its geometry. The varying range of model parameters captures maximum possible observational features. We also put light on the routinely observed Fe line emission properties based on different model parameters, ionization rate, and Fe abundances. We find that the Fe line equivalent width $W_{\rm E}$ decreases with increasing disc accretion rate and increases with the column density of the illuminated gas. Our estimated line properties are in agreement with observational signatures.
\end{abstract}

\begin{keywords}
accretion, accretion disc -- X-ray binaries:black holes -- atomic processes -- line:formation -- radiative transfer 
\end{keywords}


\section{Introduction}
X-ray observations show prominent Fe K emission lines at 6-7 ~keV and Compton hump at $\sim$20-30 keV in the spectra of black hole binaries which indicates that the material above the disc is being illuminated by the hard X-ray source. The Fe emission lines with various shapes have been routinely observed in many black hole X-ray binaries \citep[][and references therein]{Tomsicketal2014,Plantetal2014,Mondaletal2014,Debnathetal2015,Xuetal2020, Dongetal2020a} using high-resolution X-ray spectroscopy. Variations of these features with time imply that they are produced in the vicinity of the X-ray sources themselves. Several theoretical models in the literature have been proposed to explain those observed features of the emission lines. 
Detailed calculations of the radiative transfer of X-rays in an optically thick medium were carried out by \citet{Rossetal1978} and \citet{Ross1979}, solving the transfer of the continuum photons using the Fokker–Planck diffusion equation, including a modified Kompaneets operator for more realistic Compton scattering, while the transfer of lines is calculated using the escape probabilities approximation. 

In the case of black hole binaries, there are several ways to illuminate the disc by the X-ray radiation, for example, at the outer boundary of the disc, relatively less gravity, increases the height of the disc, intercepts more radiation or the winds from the corona or from the disc itself scatters the X-rays down to the disc. The effects of light bending, specifically the return of some fraction of the disc radiation to the disc itself, can become significant \citep{Cunningham1976} to enhance the strength of the emission. The emission lines generated in low mass X-ray binaries (LMXRBs) in the X-ray illuminated accretion disc was studied by \citet{KallmanWhite1989}. Authors showed that the Fe K line broadening is dominated by rotation or by Comptonization through higher optical depth, rather than from an accretion disc corona. Intensity estimation of the reprocessed emission from an irradiated slab of gas has shown that the Fe K$\alpha$ line is the strongest line in the reflection spectrum \citep{GeorgeFabian1991,Mattetal1991}. It can be by virtue of its relatively high cosmic abundance and large fluorescent yield. If the strong lines from other abundant elements such as carbon, nitrogen, and oxygen at lower energies can be detected, it would provide a lot of information on the ionisation state of the accretion disc. Several models of reflection from photoionized discs have shown that the soft X-ray features are very sensitive to the ionisation parameter of the disc material \citep{RossFabian1993}. 

Observations of Fe line emission have evidenced to be important for the determination of the spin parameter, one of the intrinsic parameters that describe black holes \citep{Laor1991, Dabrowskietal1997, Brennemanetal2006}. It was pointed out by \citet{Fabianetal1989}, that, if the reflected emission comes from the accretion disc, then relativistic and Doppler effects would broaden the emission lines, particularly in high spin black hole cases, where if reflection occurs close to the black hole \citep{Fabianetal2000,ReynoldsNowak2003,Milleretal2008,Steiner2011,Dauseretal2012}. Recent high-resolution X-ray observation indeed detected this line broadening features and was used to estimate the spin parameter of black holes \citep[][and references therein]{Iwasawaetal1996,Mondaletal2016,Dongetal2020b}.

Several models have been introduced in the literature to explain the emission lines from the disc from time to time. For example, the scattering of photons by cold electrons using Green’s functions approach was first derived by \citet{LightmanRybiki1980} and \citet{Lightmanetal1981}, and their implications for AGN observations discussed in \citet{LightmanWhite1988}. \citet{GuilbertRees1988} proposed a theoretical model to study reflected emission assuming that irradiation on the surface of the disc was weak enough so the gas remains neutral, but yet would reprocess the radiation-producing observable spectral features. \citet{Mattetal1993} studied emission properties based on mass accretion variation. \citet{Zyckietal1994} carried out similar calculations including ionisation balance and thermal balance in the medium along with the distribution of X-ray intensity with optical depth, yet neglecting the intrinsic emission inside the gas. Later, \citet{MagdziarzZdziarski1995} calculated the cold reflection using Green's function approach, highly dependent on viewing angle, however, their calculations did not include line production. Apart from the above mentioned models, there are many notable models such as, {\sc reflionx} by \citet{RossFabian1993}, {\sc TITAN} code by \citet{Dumont2000}, which was later extended by \citet{Rozanskaetal2002} to treat cases of Compton thick media. \citet{Rozanskaetal2002} demonstrated that the use of hydrostatic equilibrium is of crucial importance to study the disc illumination by the hard X-ray. All other earlier studies assumed constant density in the material. It has been argued that a plane parallel slab under hydrostatic equilibrium could represent the surface of an accretion disc more accurately \citep{Nayakshinetal2000}, and that its reflected spectrum is, in fact, different from the one predicted by constant density models \citep[see also, ][]{Rozanska1996,NayakshinKallman2001,RozanskaMadej2008,Rozanskaetal2011}. 
Recently, in the last couple of years, Garcia and his collaborators proposed a model, {\sc xillver}, \citep{Garcia2013} where authors have updated the above models adding more physical processes and atomic data, to make it more broadly applicable. However, all these models used the incident spectra from the phenomenological power law model, did not take into account the detailed accretion solution and the effects of flow parameters and its geometry.

Despite the existence of a large number of models in the literature, there is still a lack of understanding of the origin of the corona, its optical depth, and temperature profile to generate physical emission from that region, which is responsible for illuminating the disc. There are models which successfully explain the scenario of scattering of soft photons by the corona at different location of the disc \citep{SunyaevTitarchuk1980,HaardtMaraschi1993,Zdziarskietal2003}, however, without taking into account the physical origin of the corona. Therefore, a self-consistent modeling of accretion, as well as emission lines from reflected disc component, remains to be done, which motivates us to study line emissions using a physical accretion disc model. For our purpose, we use two-component solution, {\sc TCAF} \citep[][hereafter  CT95]{ChakrabartiTitarchuk1995,Chakrabarti1997}, generated theoretical spectra. According to this model, the so-called truncation of the disc is the location of the shock \citep{Chakrabarti1989}, formed by low angular momentum, hot, sub-Keplerian halo, satisfying Rankine-Hugoniot conditions.  
Numerical simulations have shown that the two-component flow forms and is stable even in presence of spatially- and temporally- varying viscosity parameters and cooling processes \citep{GiriChak2013,Giri2015,RoyChak2017}.  
Beyond the shocked region, both disc and halo accretion pile up to decide the optical depth and temperature of that region. The same region also upscatters the incoming soft radiation from the disc via inverse Comptonization. The model has indeed a reflection mechanism, where disc to shock and vice-versa interceptions are taken into account in an iterative way. The disc inclination effect has not been taken into account in the current model. Later, this model was modified to see the effects of cooling and mass loss coupled with hydrodynamics \citep{MondalChakrabart2013}. In the last couple of years, the present model is also implemented in XSPEC to study both LMXRBs \citep{Debnathetal2014,Mondaletal2014} and AGNs \citep{Nandietal2019,MondalStalin2021} data successfully. The observational studies using the current model can explain different spectral states, the evolution of quasi-periodic oscillations (QPOs), estimation of mass and spin of the black hole which covers current research topics in the field. To study the emission line we use publicly available photoionization code {\sc cloudy} C17.01 version \citep{Ferland2017}. Earlier, \citet{Bianchietal2006} used {\sc cloudy} to study the emission lines from 8 nearby Seyfert 2 galaxies and revealed that the soft X-ray emission of all the objects is likely to be dominated by the photoionized gas. Recently, \citet{Adhikari2015,Adhikari2016,Adhikari2019} used {\sc cloudy} in detail to study emission spectra in the AGN environment and its application in different geometry and thermodynamic conditions.

The goal of the present manuscript is (i) to use TCAF model generated spectra to illuminate the disc and to produce emission lines, (ii) to study how different model parameters (e.g., disc, halo accretion rates, and the disc geometry) affect line shapes and its intensity, (iii) to study the effect of Fe abundances ($A_{\rm Fe}$) and ionisation parameter ($\xi$) on the line properties in different spectral states, (iv) to estimate the profile of the equivalent width only of the Fe lines with disc mass accretion rate, (v) to see the temperature structure of the gas cloud above the disc, and (vi) to use the simulated Fe lines to compare with observed Fe line features of the black hole binaries. In this work, we are not going into the details explaining the origin of emission lines from other species than iron, which will be studied in follow-up works.

The paper is organized as follows: in \autoref{sec:model}, we discuss the details of the accretion disc model, which is used to generate the illuminating spectra and how it is used to form emission lines from {\sc cloudy}. In \autoref{sec:result}, we describe emission lines variation with different model parameters for instance, accretion rates (as the model uses two accretion components), the geometry of the flow, ionisation parameter, and Fe abundance. The equivalent width profile with disc mass accretion rate and ionisation parameter and the temperature structure of the gas cloud are also discussed. Our simulated results are also compared with observations.  In \autoref{sec:Fe_Shape}, the change in shape of Fe lines with model parameters and its intensity variation is discussed. Finally, we draw our concluding remarks in \autoref{sec:conclusion}.

\section{Modelling }\label{sec:model}
\begin{figure*}
\centering{}
\includegraphics[height=6truecm,width=12truecm,angle=0]{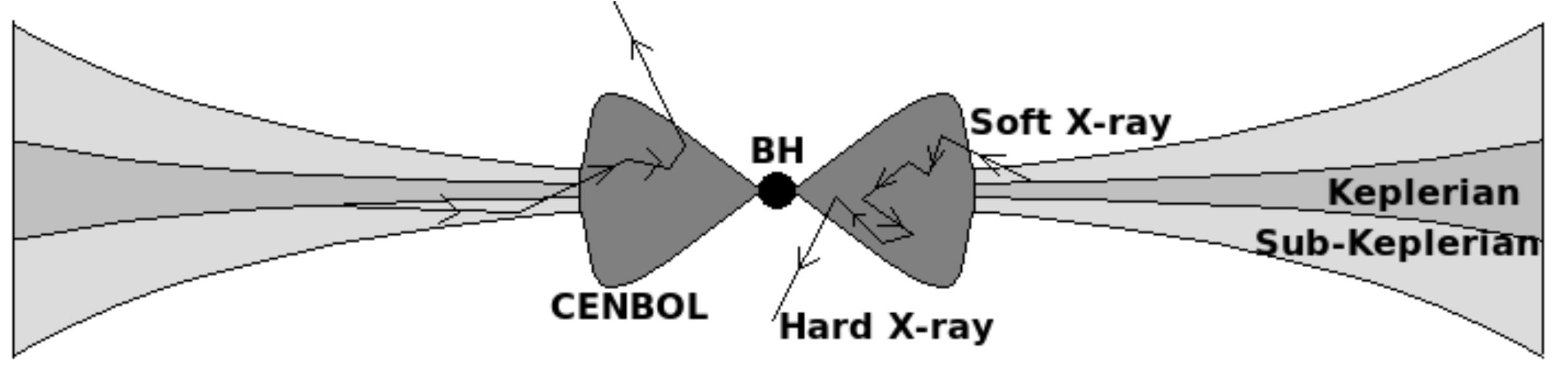}
\caption{A cartoon diagram of the geometry of the flow (CT95). The scattering of the soft photons from the Keplerian disc are the Zigzag trajectories. These photons are Comptonized by the CENtrifugal pressure supported BOundary Layer 
(post-shock region of the sub-Keplerian flow or corona) and are radiated as hard X-rays.}
\label{fig:ModelCartoon}
\end{figure*}

In order to generate emission lines, we use theoretical spectra from \citet{Chakrabarti1997} model. In \autoref{fig:ModelCartoon}, we present a cartoon diagram of the two-component model where a cold, high angular momentum Keplerian disc resides at the equatorial plane and it is flanked by the sub-Keplerian flow which we call halo, which is hot and low angular momentum flow. The optically thin pre-shock halo does not radiate efficiently therefore energy and entropy are advected with the flow. At the center of the co-ordinate, a black hole of mass $M_{\rm BH}$ is located. For the supply of soft (seed) photons for thermal Comptonization, we assume a Keplerian disc on the equatorial plane, truncated at the shock radius. This disc emits a flux of radiation the same as that produced by a \citet{ShakuraSunyaev1973} disc. The soft photons emerging out from the Keplerian disc are reprocessed via Compton or inverse Compton scattering within the corona. Injected photons may undergo a single, multiple or no scattering at all with the hot electrons in between its emergence from the Keplerian disc and its escape from the halo. As the radiation passes through the corona, the probability of repeated scattering by the same photon decreases exponentially, however, the gain in energy is exponentially higher. A balance of these two processes gives a power-law distribution of the energy density. For the temperature estimation of the corona, we solve the thermally decoupled two-temperature equations for electrons and protons. The location of the shock can be found after solving flow equations by providing initial specific energy and specific angular momentum value to the flow at the outer boundary. However, as we are not solving the transonic flow coupled with the radiative transfer, we use corona size or shock location as a free input parameter of the model.

In the present model, we have computed the model spectra due to thermal Comptonization only. As the innermost region of the disc is rapidly falling in and is, in fact, supersonic \citep{Chakrabarti1989,ChakrabartiTitarchuk1995,MondalChakrabart2013}. At high accretion rates, when the flow becomes cold and the thermal Comptonization can be ignored, the bulk motion takes the role of energy shifting of seed photons. This so-called bulk motion Comptonization (BMC) spectrum is decided by the upper limit of the velocity of the infalling matter and thus the spectral slope saturates to around 2.0 even when the accretion rates are varied. The theoretical works \citep{ChakrabartiTitarchuk1995,TitarchukZannias1998}, Monte Carlo simulations \citep{LaurentTitarchuk1999}, and the observational results \citep{Shaposhnikovetal2009,TitarchukSeifina2009} all  point  to  these  saturation  effects.  Since  this  property  is  solely due to the unique properties of a black hole accretion, this is not affected by our analysis of thermal Comptonization, valid for relatively lower accretion rates of the Keplerian component. 

\begin{figure} 
\centering
\includegraphics[width=8.0truecm,trim={0cm 0.0cm 0cm 0.0cm}, clip]{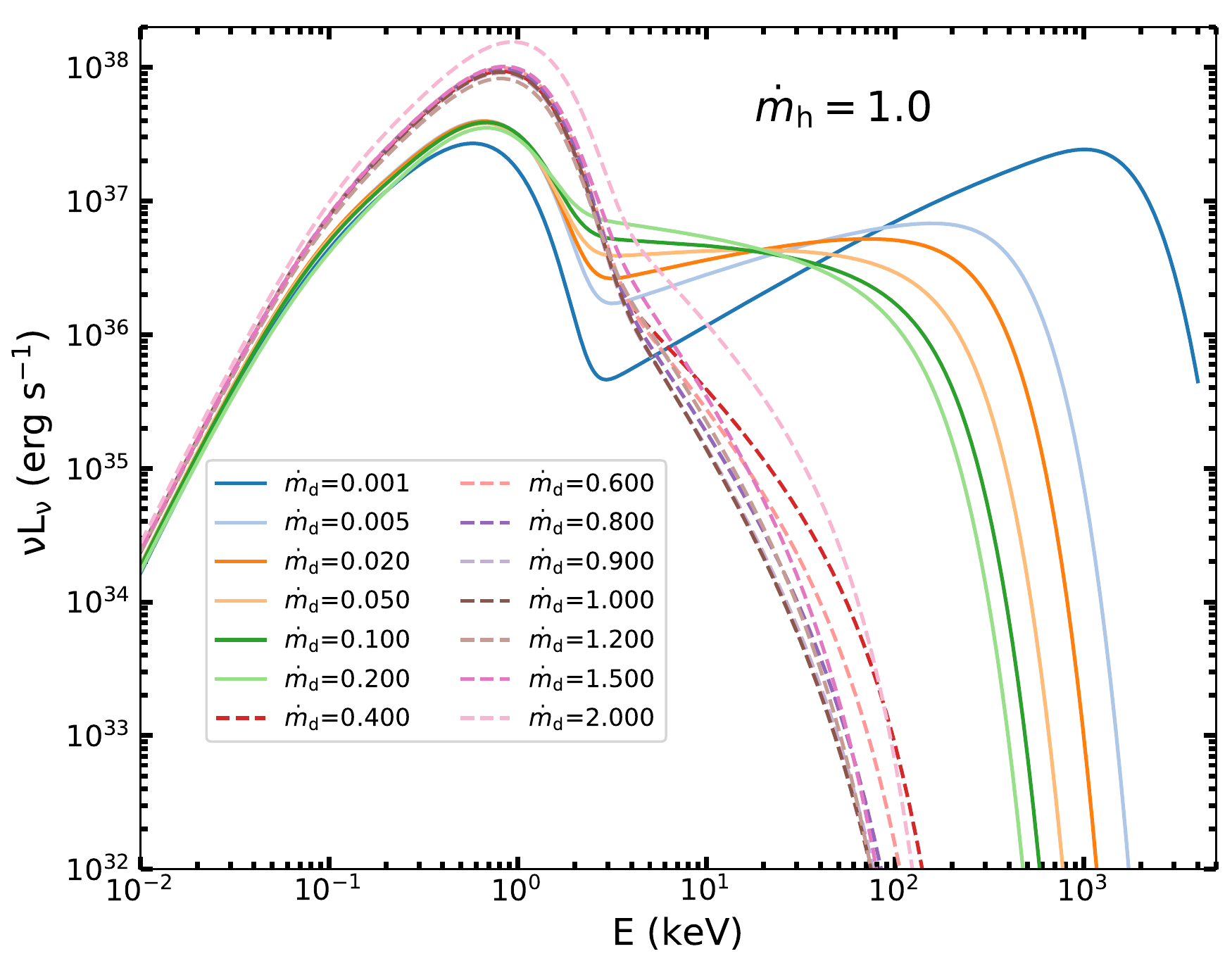}
\caption{Incident radiation spectra derived from the model for various values of mass accretion rates, $\dot{m_{\rm d}}$, when the other model parameters are fixed at $\dot{m_{\rm h}}=1.0$, $X_{\rm s}$=30, and $R$=2.5. Two clusters of SEDs: a)first cluster with cut off around 100 keV shown by dotted colored lines, and b) second cluster which with cut off around 1000 keV shown by solid colored lines. The detail parameters for other SEDs are mentioned with the corresponding figures. }
\label{fig:Incident_SEDs}
\end{figure}

The theoretical TCAF model uses five parameters namely, (i) mass of black hole (M$_{\rm BH}$ in $M_\odot$ unit), (ii) disc mass accretion rate ($\dot m_{\rm d}$), (iii) halo mass accretion rate ($\dot m_{\rm h}$). The accretion rates are in the Eddington accretion rate  ($\dot M_{\rm Edd}$) unit, (iv) location of the shock or boundary layer of the corona ($X_{\rm s}$ in $r_{\rm s}$ =$2 GM_{\rm BH}$/$c^2$ unit, where, $G$ and $c$ are the gravitational constant and speed of light respectively). Here, $X_{\rm s}$ is the boundary layer of the corona and we are solving steady state equations, therefore at any instant standing shock will form at a particular location,  and (v) shock compression ratio ($R=\rho_- / \rho_+$, where, $\rho$ is the density of the pre (-) and post (+) shock flow). Here, we briefly discuss how model parameters change the shape of the emitted spectra. Parameters (ii) to (v) collectively defines the properties of the disc and corona of the flow, for instance, the electron density and temperature, photon spectrum, and density. The fractional interception of soft photons from the disc by the corona and vice-versa are taken into account. During the scattering of soft photons by the corona, cooling also takes place, which is also responsible for changing spectral shapes \citep{Doneetal2007,MondalChakrabart2013,YuanNarayan2014}. All of these depend on the mass of the black hole, thus mass estimation can be done satisfactorily \citep{Mollaetal2017} by analyzing full outburst data of different X-ray missions, by keeping the mass of the BH as a free parameter or by keeping the normalization fixed for a particular outburst for a particular BH, details are explained in the paper. According to \citet{Chakrabarti1997}, if one increases the halo rate keeping other parameters unchanged, the model will produce a hard spectrum. Similarly, increasing the disc rate leaving other parameters frozen will produce a soft spectrum. \autoref{fig:Incident_SEDs} shows the typical model spectra for varying $\dot m_{\rm d}$ when other parameters are fixed. Increasing $\dot m_{\rm d}$ generates softer spectra. The parameter values for each spectrum are given in the figure. The solid lines show the set of spectra with cutoff at higher energy $\sim 1$ MeV whereas the dotted lines show cutoff around 100 keV. When the location of the shock is increased keeping other parameters fixed, the spectrum will become harder. A similar effect can be also seen for the compression ratio (R). The effects are discussed later in \autoref{subsection:geometry_vary}. In general, in outbursts of black hole binaries, during rising and declining phases with different spectral states, all the parameters change smoothly in a multidimensional space \citep[][and others]{Mondaletal2014,Debnathetal2015,Chatterjeetal2016,Janaetal2016}. The accretion rate ratio (ARR, the ratio of disc rate with halo accretion rate) is also an indicator of spectral states change \citep{Mondaletal2014}, where the spectra move from hard (HS) to soft state (SS) through intermediate states when ARR changes from low to high.

To generate the model spectra we use the outer boundary of the disc at 500~$r_{\rm s}$, extend up to $X_{\rm s}$ and corona starts from $X_{\rm s}$, and extend up to 3~$r_{\rm s}$. The illuminating radiation contains different components, including blackbody, Comptonization, and reflection. Photo-ionisation processes are simulated with numerical spectral synthesis code {\sc cloudy} version C17.01 \citep{Ferland2017}. The {\sc cloudy} solves different radiation steps to generate emission lines \citep[for detail code description and most recent updates see][]{Ferlandetal1995,Ferland2017}.

In {\sc cloudy}, 
a slab of gas is divided into a large number of thin zones which results a plane parallel open geometry. In this geometry, the photon once escaped will be lost to infinity without further interactions. A spherical closed geometry can be implemented to include the multiple interactions of an escaped photon as well. However, for this case of studying illumination of disc by the photons from corona, the assumption of plane parallel geometry is more realistic. Here, we note that the radiative transfer is solved in 1 dimension only. The radiation field from the accretion disc is normally incident on the slab of gas and its reprocessing is initiated. We do not take into account the incidence at different angles. The gas density which has been used as the input parameter is taken from the mass accretion rate, and the column density ($N_{\rm H}$) is kept constant to $1.28\times 10^{21}$ cm$^{-2}$. However, the effect of $N_{\rm H}$ has been verified in our study in a range. For the detail discussions on the different possible geometries of the radiation matter interactions and their implementation in various astrophysical environments, we refer the reader to the {\it Hazy1\footnote{www.nublado.org}} documentation of {\sc cloudy}. Moreover, \citet{Adhikari2015,Adhikari2019b} have extensively discussed the use of open and closed geometries in {\sc cloudy} for various cases of AGN absorption and emission regions.

The next important assumption we employ in {\sc cloudy} modelling is the use of {\it constant density} case. This means that the gas number density is kept constant to a given value across all the zones of the given gas cloud. Here we note that,
an alternative situation of {\it constant pressure} can arise when a gas cloud is illuminated by the radiation energy where the gas number density is stratified across the zones of cloud. This phenomenon of radiation pressure confinement is discussed and implemented in various photoionized environments of different astrophysical systems in the literatures \citep{Rozanska2006,Stern2014,Baskin2014,Adhikari2015,Adhikari2019,Adhikari2019b}. However, \citet{Adhikari2018} have shown that when there is high gas number density and hence the high gas pressure, the radiation pressure confinement is very weak, and the model structures are very similar between {\it constant density} and {\it constant pressure} assumptions. Since the gas number density in the accretion disc of black hole X-ray binaries are quite high, we choose to use the {\it constant density} assumption for simulating the photoionisation process.

The level of ionisation is determined by balancing all ionisation and recombination processes. Ionisation processes include photo, Auger, collisional ionisation, and charge transfer. Recombination processes include radiative, low-temperature dielectronic, high temperature dielectronic, three-body recombination, and charge transfer. The free electrons are assumed to have a predominantly Maxwellian velocity distribution with a kinetic temperature determined by the balance between heating (photoelectric, mechanical, cosmic-ray, etc.) and cooling (predominantly inelastic collisions between electrons and other particles) processes. The line emission and continuum radiative transfer processes are solved simultaneously.

The incident radiation illuminates the disc and ionizes it. The ionisation parameter is calculated using the equation:
\begin{equation}
\xi=\frac{4\pi F_{\rm inc}}{n_{\rm H}}  \text{erg cm s$^{-1}$},
\end{equation}
where, $F_{\rm inc}$ is the Hydrogen ionizing flux integrated in the range 1-1000 Rydbergs and $n_{\rm H}$ is Hydrogen number density in cm$^{-3}$ unit. The chemical composition of the disc is set to default {\it Solar} values in {\sc cloudy} and it is mentioned when the Fe abundances ($A_{\rm Fe}$) is varied. {\it Solar} abundances in {\sc cloudy} are adopted from \citet{Grevesse1998}.
The electron number density at some radius r of the flow is calculated from $\dot m_{\rm h}$ above the disc, given by, $n_{\rm e}=\frac{\dot m_{\rm h}}{4\pi r^2 m_{\rm p} \varv}$ in $\text{cm}^{-3}$ unit, where $m_{\rm p}$ and $\varv$ are the mass of the proton and velocity of the inflow, which can be written as $1/r^{1/2}$, when the disc radius is away from the black hole. Thus the typical value of $n_e$ varies in a range from $1.4 \times 10^{10}$-$1.4 \times 10^{13} \text{cm}^{-3}$ for the accretion rate range 0.001-1 $\dot M_{\rm Edd}$ and $r$=500$r_{\rm s}$. For the purpose of this work, the electron density self consistently derived from our model can be used as the hydrogen number density $n_{\rm H}$ in the cloudy computations. The model parameters used to construct the spectral radiation shape or spectral energy distribution (SEDs), which are used in the {\sc cloudy} modelling are discussed in each figure.

\section{Results} \label{sec:result}
For the gas cloud above the disc which is illuminated by the central radiation, we have calculated model reflection spectra covering a wide range of disc parameters, with disc rate ($\dot m_{\rm d}$=0.001 - 2.0), the halo rate ($\dot m_{\rm h}$=0.001, 0.01, 0.1, and 1.5), the shock location ($X_{\rm s}$=10, 30, 50, 80, and 100), the shock compression ratio ($R$=1.5, 2.5, 3.5, and 4.0), the ionisation parameter ($\log \xi$=1, 2, 3, and 4), and the iron abundance ($A_{\rm Fe}$ = 0.5, 1.0, 2.0, and 5.0) relative to its Solar value. The mass of the black hole is fixed at $7 M_\odot$ throughout the paper. These range of parameter values take into account for most of the observed spectral states and features of black hole binaries, and all parameters have significant effect on emission lines. For simplicity, abundances of all other elements considered are kept fixed at Solar values.

\subsection{The effect of varying $\dot m_{\rm d}$}
\begin{figure} 
\centering
\includegraphics[height=8.0truecm]{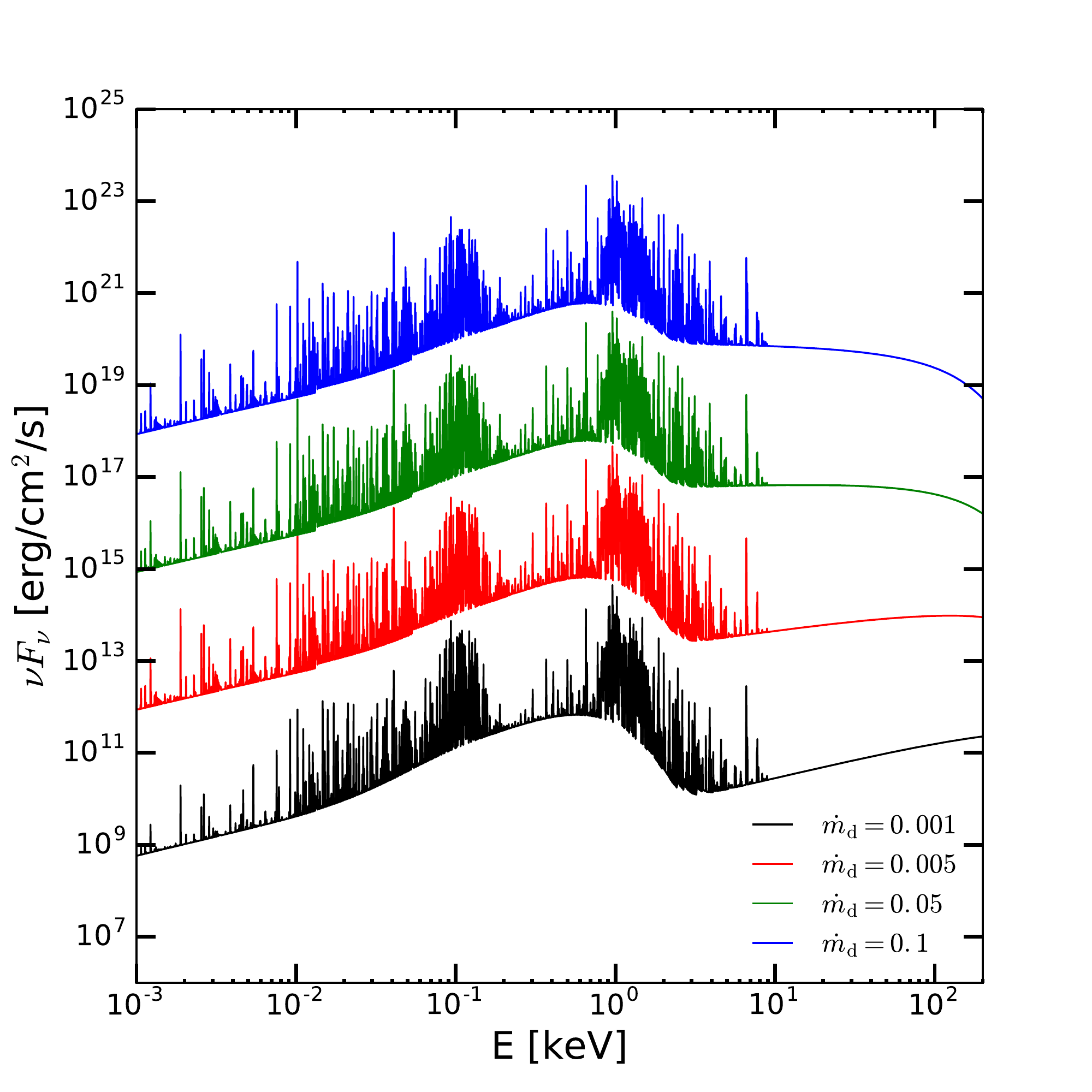}
\caption{Reflected spectra for four different values of the $\dot m_{\rm d}$, with $\dot m_{\rm d}$ = 0.001, 0.005, 0.05, and 0.1. The incident spectrum has $\log \xi$ = 3, and iron has Solar abundance, the other theoretical model parameters are $\dot m_{\rm h}$ =1.0, $M_{\rm BH}$=7.0~M$_\odot$, $X_{\rm s}$=30, and $R=2.5$. Successive spectra have been offset by factors of 1000 for clearer visibility.}
\label{fig:reflection_md}
\end{figure}

\autoref{fig:reflection_md} shows reflected spectra for four different values of $\dot m_{\rm d}$, mentioned in the figure. In all cases, the incident spectra have $\dot m_{\rm h}$=1.0, $X_{\rm s}=30$, $R=2.5$, $\log \xi=3$, and the iron has Solar abundance ($A_{\rm Fe}$ = 1.0). Increasing $\dot m_{\rm d}$ increases the number of soft photons, thus the intensity of emission lines at the high energy regime of the spectrum decreases, might be due to the presence of less number of high energy photons, however, the possibility of forming more lines is observed. It is clear from the emission spectra that both illuminating flux and the spectral shape of the ionizing radiation incident on the surface of the disc have a significant impact on the ionisation balance of the gas, and thus on the reflected spectrum. 

The equivalent width $W_{\rm E}$ of an emission line is calculated by using an expression,
\begin{equation}
   W_{E}=\int_{5.5 \text{keV}}^{7.2 \text{keV}} \frac{F_{E}^{C}-F_{E}^{t}}{F_{E}^{C}} dE,
\end{equation}
 where $F_{E}^C$ and $F_{E}^t$ are the fluxes in the reflected continuum and around the line energy centroid respectively.
  For the purpose of studying the emission from highly ionized Fe, we sum up the $W_{\rm E}$ of Fe-emission lines in the energy band $5.5 - 7.2$ keV and define it as the Fe-line $W_{\rm E}$. In the computations of the $W_{\rm E}$ in this energy band, the contribution of the ions Fe {\sc ii} to Fe {\sc xxvi} are considered. 

\begin{figure} 
    \centering{
    \includegraphics[height=6.0truecm]{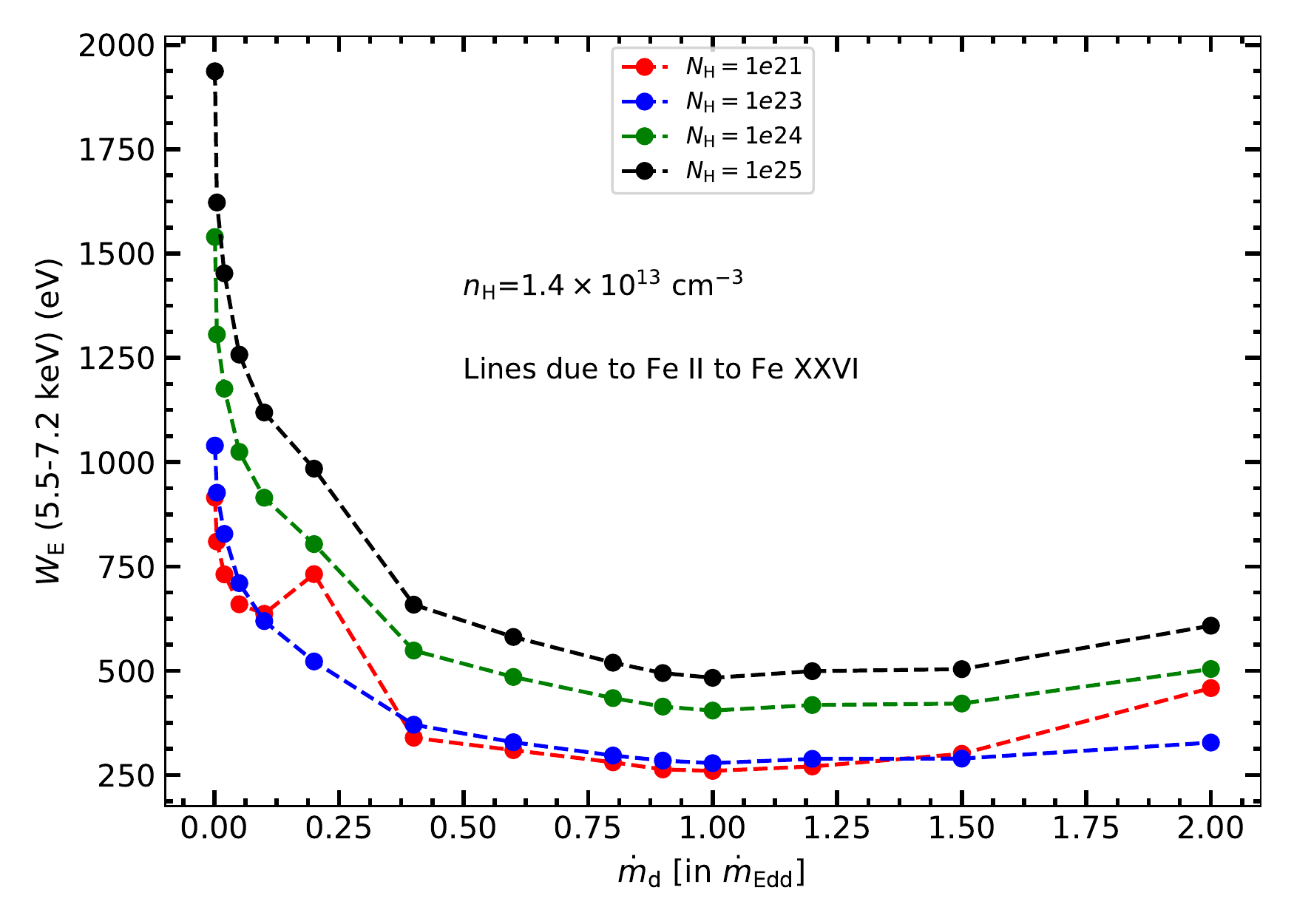}}
     \caption{$W_{\rm E}$ of Fe line integrated in the energy range $5.5-7.2$ keV as a function of $\dot m_{d}$. We extend $\dot m_{d}$ values to super-Eddington rate to see the trend of behavior. Various colours represent the $W_{\rm E}$ computed for various values of $N_{\rm H}$. Other theoretical parameters used in these computations are: $\dot m_{\rm h}$ =1.0,  $M_{\rm BH}$=7.0~M$_\odot$, $X_{\rm s} = 30$ and R = 2.5 respectively. In the {\sc cloudy} modelling, $\log \xi$ = 3.0 and Solar values for Fe abundances are used. Various colors in the plot depicts the varying $N_{\rm H}$ employed.
     }
    \label{fig:ew}
\end{figure}

\autoref{fig:ew} shows that the general trend of $W_{\rm E}$ is decreasing with increasing $\dot m_{\rm d}$. This is expected as at low accretion rate most of the photons are concentrated at the high energy part of the spectrum, where the Compton scattering becomes important, therefore hard spectra are more efficient in heating the illuminated layer, which emits more lines. As the accretion rate increases, the incident spectrum becomes softer thus less number of hard photons participate in line emission. It also shows that increase in column density increases $W_{\rm E}$, which satisfies that $W_{\rm E}$ $\propto$ $N_{\rm H}$. The red, blue, green, and black lines correspond to $N_{\rm H}$ values $10^{21}$, $10^{23}$, $10^{24}$, and $10^{25}$ respectively. The $W_{\rm E}$ is estimated for the lines Fe{\sc ii}-Fe{\sc xxvi}. Our estimated variation of $W_{\rm E}$ is in the same line of \citet{ZyckiBozena1994,Garcia2013}, where authors showed that $W_{\rm E}$ decreases with increasing photon index ($\Gamma$), which is the same in our model as increase in $\dot m_{\rm d}$ makes the spectrum softer thus the higher value of ($\Gamma$). A recent analysis of MAXI~J1820+070 \citep{Xuetal2020}, showed a similar agreement. However, opposite behavior can also be observed if the lines originate from the inner edge of the disc, where the disc moves inward with accretion rate, which increases rotational velocity, therefore increases line broadening, thus the $W_{\rm E}$ \citep{Tomsicketal2009,Debnathetal2015}. The $W_{\rm E}$ is also an important measurable quantity to diagnose the evolution of QPOs in the accretion disc, reported in Galactic black hole GRS~1915+105 \citep{Millerhoman2005}. Earlier the current model was also used to show the evolution of QPOs with disc accretion rate for the outbursting candidates \citep{Mondaletal2014,Chakrabartietal2015}, where the QPO frequency increases with accretion rate. Thus combining these two effects in the future we will be able to study the QPOs evolution consistently from the Fe line fitting. Apart from the variation of the Fe line width, our results also include emission lines at low energies e.g., N and O to a range between 0.4 to 0.8 keV, which have been observed in high-resolution {\it XMM-Newton} observation of black hole candidate Swift~J1753.5-0127 \citep{Mostafaetal2013}. The range of our simulated $W_{\rm E}$ agrees with estimations from the model fitted observed data \citep{TitarchukSeifina2009}. However, it should be noted that as the flow comes closer to the black hole with increasing disc accretion rate, BMC becomes effective, and also the gravitational effects take a significant role, that can skew the line properties \citep{TitarchukSeifina2009}.

\subsection{The effect of varying $\dot m_{\rm h}$}
\begin{figure} 
    \centering{
    \includegraphics[height=8.0truecm]{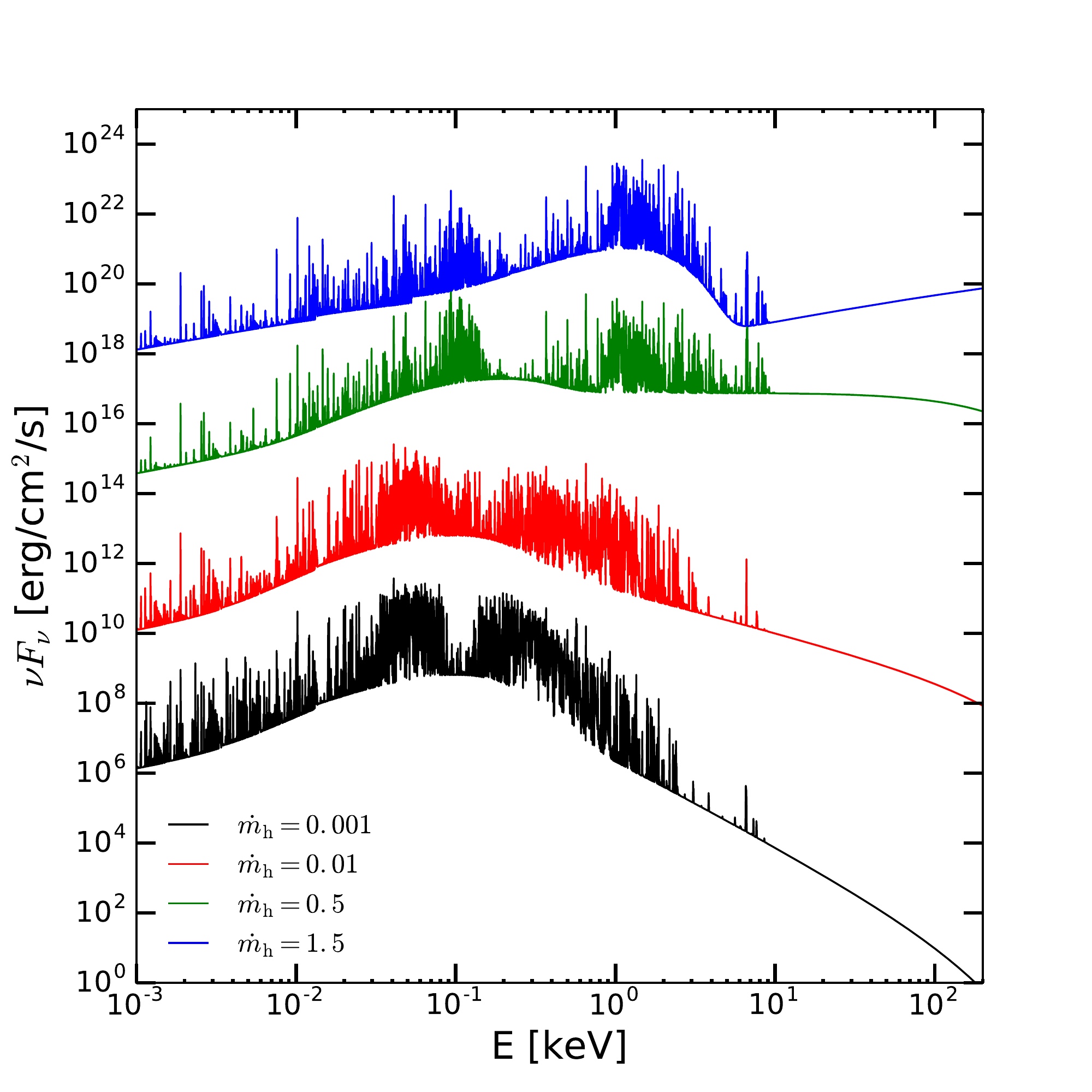}}
    \caption{Reflected spectra for four values of the $\dot m_{\rm h}$, with $\dot m_{\rm h}$ = 0.001, 0.01, 0.1, and 1.5. The incident spectrum has $\log \xi$ = 3, and iron has Solar abundance, the other theoretical model parameters are $\dot m_{\rm d}$ =0.02, $M_{\rm BH}$ = 7.0~M$_\odot$, $X_{\rm s}$ = 30.0, and R = 2.5. Successive spectra have been offset by factors of 1000 for clearer visibility.}
    \label{fig:reflection_mh}
\end{figure}

In \autoref{fig:reflection_mh}, we carry out the same task for four different values of $\dot m_{\rm h}$, when $\dot m_{\rm d}$=0.02 and keeping all other parameters the same as in \autoref{fig:reflection_md}. As we discussed for increasing $\dot m_{\rm h}$ spectra become harder thus increases the high energy photons at high energies which affect the emission features, as the harder illuminating spectra have greater ionizing efficiency, change the shape of the reflection spectra significantly. For $\dot m_{\rm h}$=0.1 and 1.5 (HS), the illuminated gas is more highly ionized than for $\dot m_{\rm h}$=0.001 (SS) and 0.01 (intermediate state). One can observe that more emission lines are visible in the high energy regime and less lines are observed at low energy regime which is due to the outer layer of the illuminated cloud gets fully ionized for higher $\dot m_{\rm h}$. However, an opposite feature is observed for the low $\dot m_{\rm h}$ spectra. In the intermediate state (red curve), throughout the energy band emission lines are clearly visible with a relatively high intensities. 

\subsection{The effect of varying geometry} \label{subsection:geometry_vary}
\begin{figure*} 
    \centering{
    \includegraphics[height=8.0truecm]{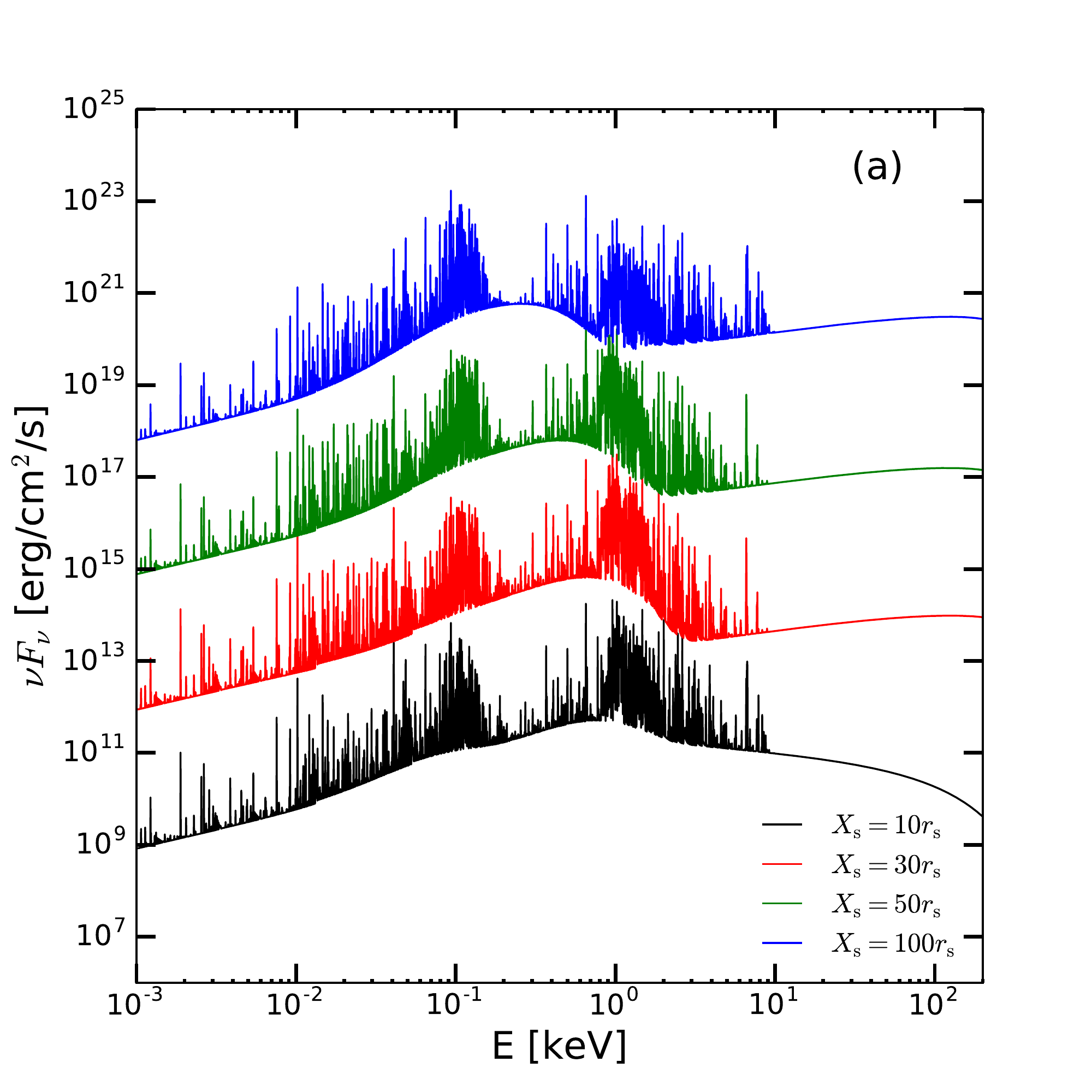}
    \hspace{-0.5cm}
    \includegraphics[height=8.0truecm]{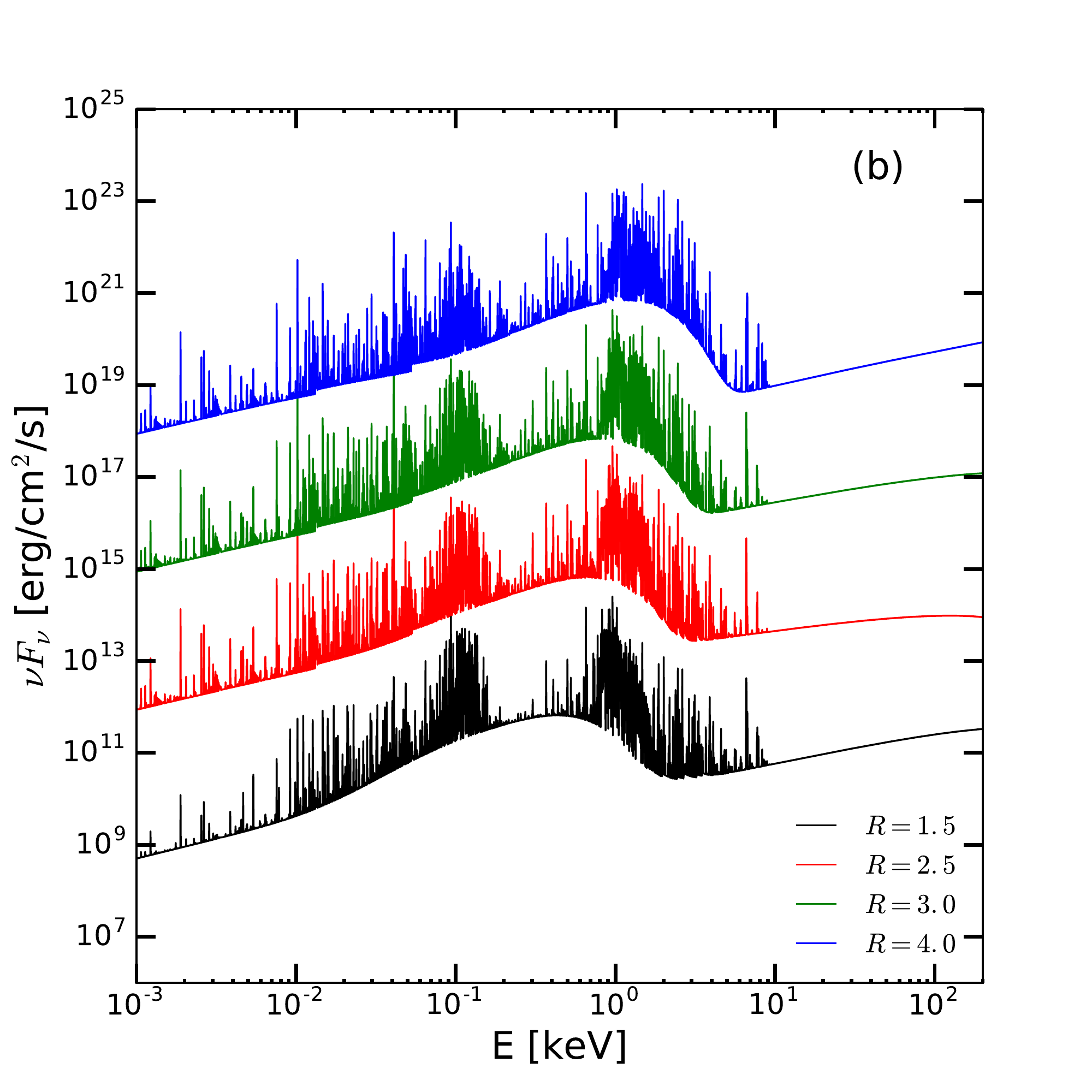}}
    \caption{Reflected spectra (a) for four values of the $X_{\rm s}$, with $X_{\rm s}$ = 10, 30, 50, and 100, when $R$=2.5 and (b) for four values of R, with $R$= 1.5, 2.5, 3.0, and 4.0, when $X_{\rm s}$=30. The all incident spectra have $\log \xi$ = 3, and iron has Solar abundance, the other theoretical model parameters are $\dot m_{\rm d}$ =0.005, $\dot m_{\rm h}$ = 1.0, and $M_{\rm BH}$=7.0~M$_\odot$.}
    \label{fig:reflection_xs}
\end{figure*}

\autoref{fig:reflection_xs} shows emission lines due to variation in accretion geometry e.g. varying size of the corona and its height (h$_s$). Here, we choose four different values of corona size i.e. $X_{\rm s}$, with $X_{\rm s}$=10, 30, 50, and 100, and four different values of R, with $R$=1.5, 2.5, 3.0, and 4.0, from SS to HS through intermediate states. Both parameters change h$_s$ and also the optical depth of the corona, thus the emitted spectra. The change in $X_{\rm s}$ and $R$, changes the h$_s$ as $\sim$ $\sqrt{(R-1)}X_{\rm s}/R$ and the total accretion rate at the post-shock region after pilling up of matter can be written as $R \dot m_{\rm h}$+$\dot m_{\rm d}$ \citep{Mondaletal2014b}, indicates that the physical quantities (temperature and optical depth etc.) of the corona depends on $R$. Furthermore, varying $R$ during the outburst phase also means variable mass outflow, as the TCAF model proposed that the corona is the base of the mass outflow/jet, which extracts thermal energy from the inner hot region and makes the spectrum softer \citep{Chakrabarti99,Singhskc11,Mondaletal2014b}, therefore change in emission line properties. However, in the present modeling we do not include jet effects. \autoref{fig:reflection_xs}a shows the lines for different $X_{\rm s}$, at the intermediate $X_{\rm s}$, the intensity of a few lines is higher than low and high $X_{\rm s}$, some extra lines are visible around 0.3~keV. \autoref{fig:reflection_xs}b, shows the lines for different values of R, with increasing R, spectrum moves from SS to HS, thus high energy photon intensity increases, which increases the ionisation rate. Therefore more lines are visible with increasing R, especially below 0.01~keV and between 0.2-0.4~keV.

\subsection{The effect of varying $\xi$} \label{subsec:varyxi}
\begin{figure*} 
    \centering{
    \includegraphics[height=8.0truecm]{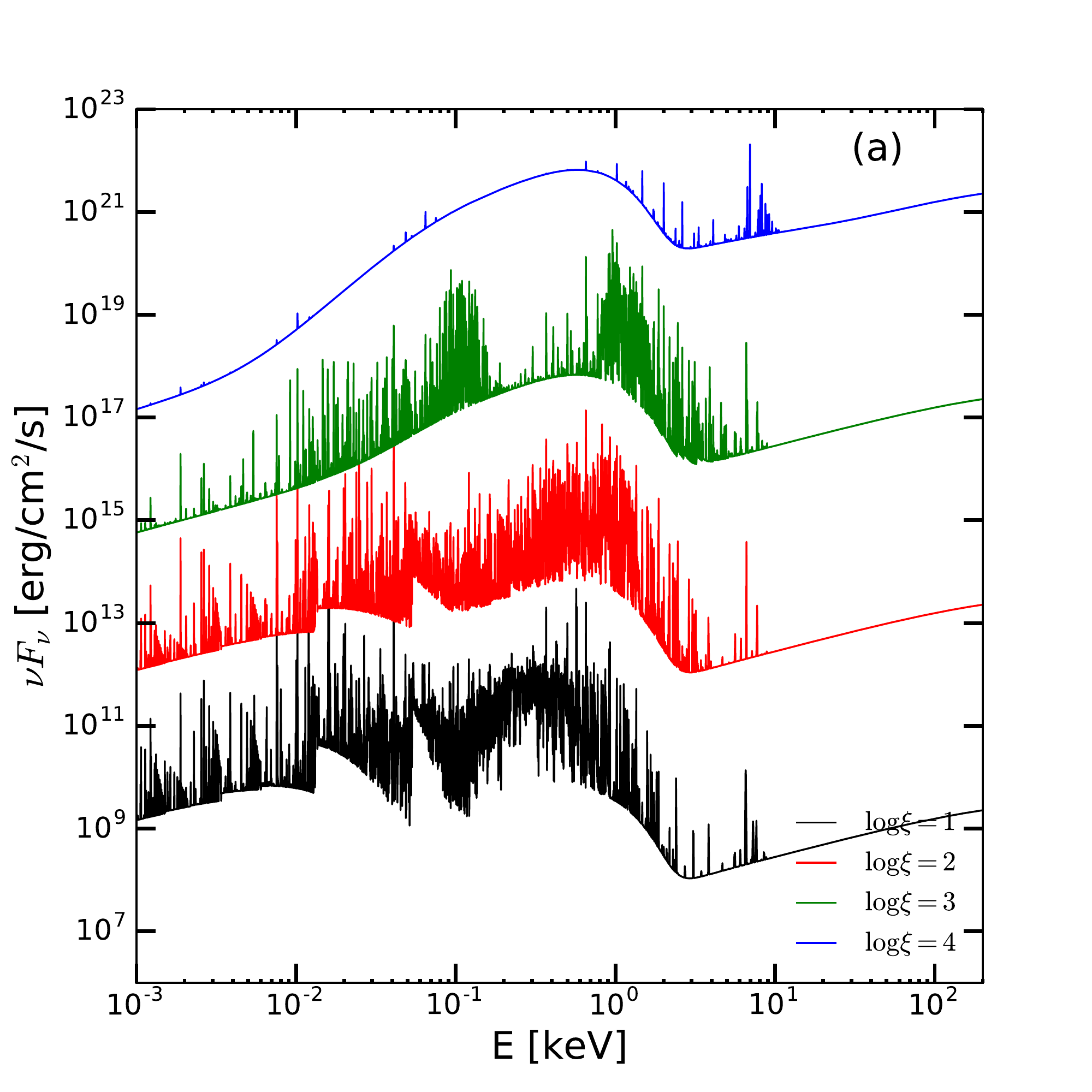}
    \hspace{-0.5cm}
    \includegraphics[height=8.0truecm]{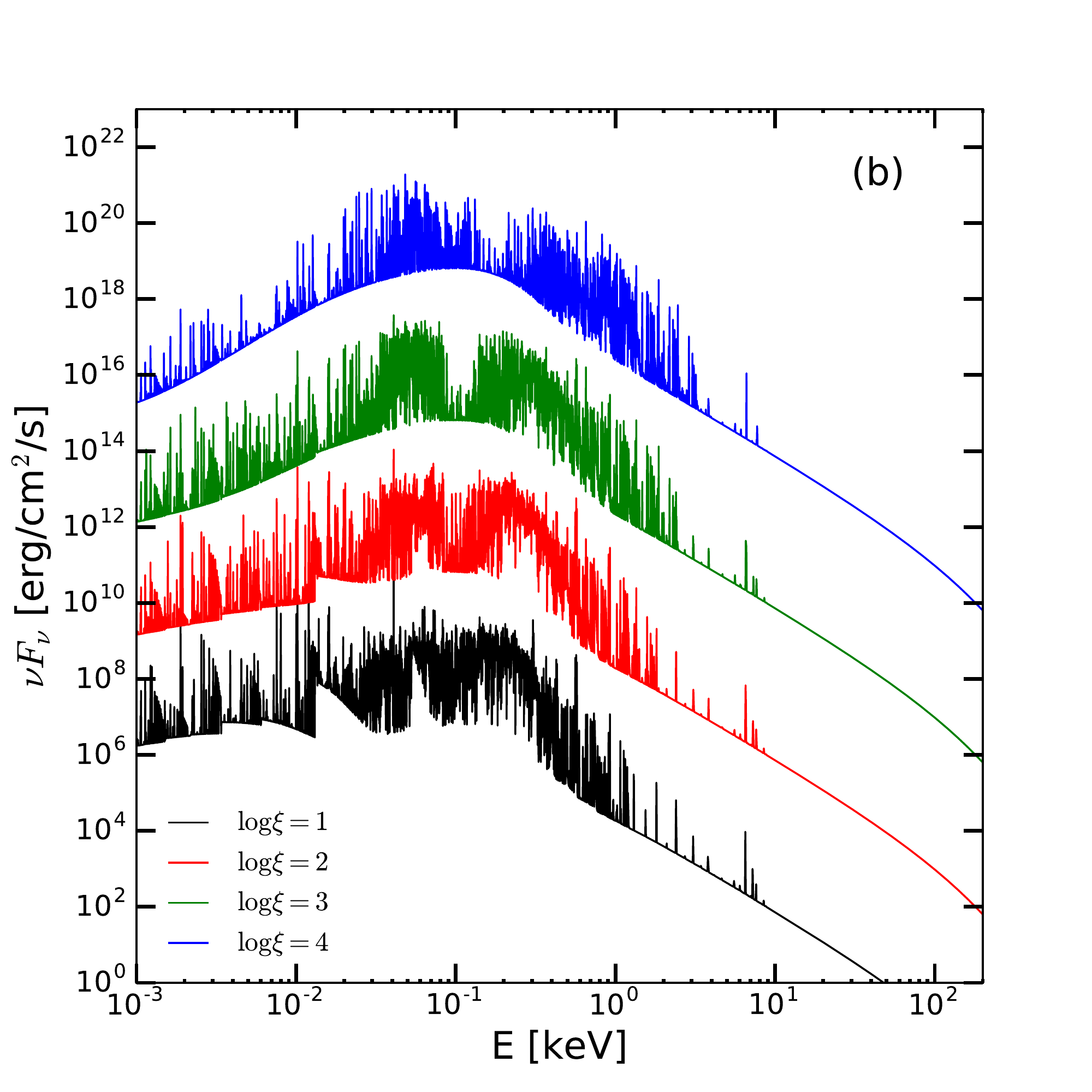}}
    \caption{Reflected spectra for four values of the $\log \xi$, with $\log \xi$ = 1, 2, 3, and 4. The iron has Solar abundance, $M_{\rm BH}$=7.0~M$_\odot$, $X_{\rm s}$ = 30 and R = 2.5. The other disc parameters are (a) $\dot m_{\rm d}$ =0.001, $\dot m_{\rm h}$ = 1.0, and (b) $\dot m_{\rm d}$ =0.02, $\dot m_{\rm h}$ = 0.001. Successive spectra have been offset by factors of 1000 for clearer visibility.}
    \label{fig:reflection_xiAfe}
\end{figure*}

The flux of the continuum that hits the Cloudy gas is fixed by the gas density and the ionization parameter as described in Eq.~1. The open geometry (a plane parallel slab with several thin zones) of the emitting gas in Cloudy is assumed which is different from the geometry of how TCAF is formulated. For a given gas cloud, the ionization parameter at the illuminated surface is defined. However, in the inner zones of the slab, it can change according to Eq.~1. Nevertheless, with our assumption of constant density gas (with thin slab) this change is negligible. For a closed (spherical) geometry, the change in ionization parameter can be significant and should be represented by a distribution \citep{Adhikari2015}. Also, if one assumes a constant pressure gas (instead of constant density), then, there can be several orders of magnitude difference in ionization parameter between the illuminated side and the back side of the slab \citep[][and references therein]{Adhikari2016,Adhikari2018,Adhikari2019}. Therefore, we examine the effect of $\xi$ on the emission lines for different values of $\xi$ applied to two different values of ARR, keeping the other TCAF model parameters fixed. In both panels of \autoref{fig:reflection_xiAfe}, each curve corresponds to a particular value of log$\xi$=1, 2, 3, and 4. The other model parameters are $M_{\rm BH}$=7.0~M$_\odot$, $X_{\rm s}$ = 30, R = 2.5, and the iron has Solar abundance. In each panel, the plotted spectra have been rescaled for visual clarity. The scaling factors are, from bottom to top, 1, 10$^3$, 10$^6$, and 10$^9$. The left panel in \autoref{fig:reflection_xiAfe}a shows emission spectra when ARR=0.001, i.e. a very hard state of the illuminating spectrum. On the other hand, \autoref{fig:reflection_xiAfe}b, when ARR=20, i.e. a softer state of the illuminating spectrum, and the effects are indeed visible. For the same ARR, increasing $\xi$ means a higher ionisation rate, which raises the temperature of the illuminated region of the slab thus ionize the gas at a larger optical depth. Hence ions from lower atomic number (Z) elements are completely stripped from all their elements, while heavy Z elements get partially ionized. Thus the emission spectra lack lines from most of the low-Z elements, are either absent or the intensity of the lines went down abruptly when $\log \xi$ increases and progressively showing for heavy-Z elements. This specific trend is observed for $\log \xi$=4. The Fe K$\alpha$ line is peaking at ~7keV. However, an opposite scenario is observed in \autoref{fig:reflection_xiAfe}b when both ARR and $\xi$ are high. All lines are visible even for higher $\xi$ in comparison of low ARR (hard state) in the left panel. This infers that in the hard state, the partly ionized disc can emit more lines as more hard photons take part in line emission. However, for a fully ionized disc with harder illumination, it emits fewer lines. Whereas in the soft state, the disc emits more lines even for a higher ionisation parameter ($\xi$) as the disc does not get completely ionised. However, the left panel with $\log \xi$=4 can show up emission lines at low energies if the column density of the gas cloud is increased. If the slab thickness is less and the radiation is hard, it will pass through the slab with little scattering or without interacting with the medium.

\begin{figure} 
    \centering{
    \includegraphics[width=8cm]{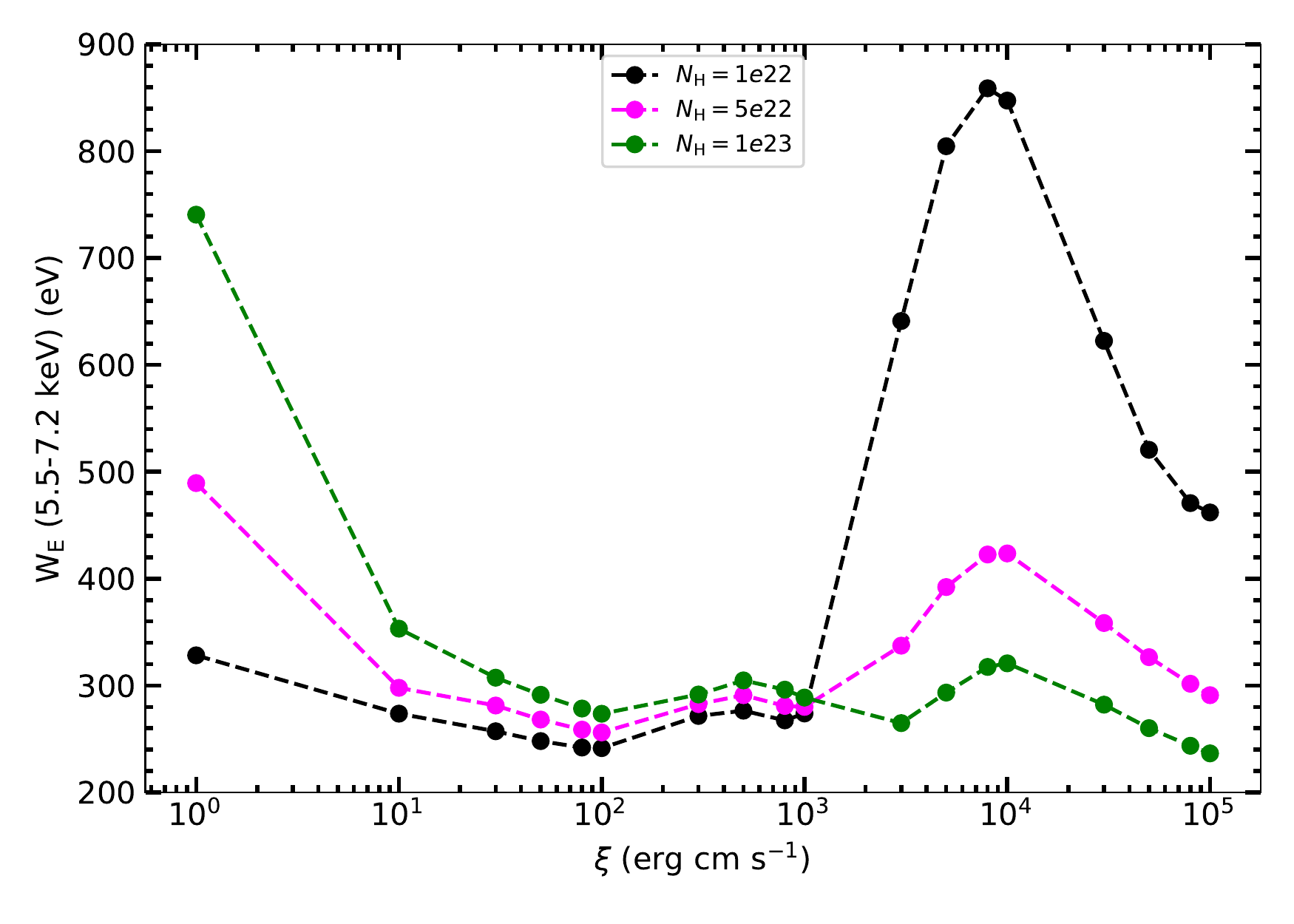}}
    \caption{Variation of equivalent widths as a function of ionisation parameter $\xi$. The incident SED used for this case is parametrized by: $\dot m_{\rm d}$ =1.2, $\dot m_{\rm h}$ =1.0,  $M_{\rm BH}$=7.0~M$_\odot$, $X_{\rm s}$ = 30 and R = 2.5 respectively. In the {\sc cloudy} modelling, Solar abundances are used and various colors in the plot depicts the varying $N_{\rm H}$ employed.}
    \label{fig:xi_ew_vary}
\end{figure}

\autoref{fig:xi_ew_vary} shows the plot of $W_{\rm E}$ versus ionisation parameter, where $W_{\rm E}$ tends to decrease with increase in ionisation parameter until low ionisation region, however, it sharply increases in higher ionisation regime, $\xi > 10^3$. After some value of ionisation, $\xi \sim 10^4$, again the $W_{\rm E}$ decrease which is because of a wider part of the disc becomes fully ionized, these typical behaviors can also be seen mainly in case of Seyfert~1 galaxies atmosphere \citep{ZyckiBozena1994}. As evidenced from \citet{ZyckiBozena1994} $W_{\rm E}$ becomes maximum around $\xi=3\times 10^{8}$ for Seyfert~1 galaxies and decreases for further increase in $\xi$. The difference of $\sim$ 4 orders of magnitude in the $\xi$ value at which $W_{\rm E}$ peaks in X-ray binaries and Seyfert galaxies is due to the considerable difference in their luminosity. Considering this point, the results from Figure~\ref{fig:xi_ew_vary} of this paper corroborate with the results of \citet{ZyckiBozena1994} (Figure 19 of their paper). The black, pink, and green curves corresponds to different values of $N_{\rm H}$. The peak value of $W_{\rm E}$ is highest for the lowest value of $N_{\rm H} = 10^{22}$. This is due to the fact that, the larger fraction of the slab of gas with low column density is at high temperature, and thus more ionized, as compared to that with higher column density. This fact can also be realized by looking at the Figure~\ref{fig:od}, where the temperature for the lowest column density stays high even at the backside of the slab. However, for the case with highest column density, the temperature falls off rapidly with the depth of the slab and unable to ionize the Fe inside. Here, we study the dependence of our result on the ionization parameter defined at the surface of the gas cloud. From observations, $\xi$ is always constrained to have a value in the range. This is why we are exploring the range of values. Similar studies were also carried out by \cite{ZyckiBozena1994} and \cite{Garcia2013}.

\subsection{The effect of varying $A_{\rm Fe}$}
\begin{figure*}
    \centering{
    \includegraphics[height=8.0truecm]{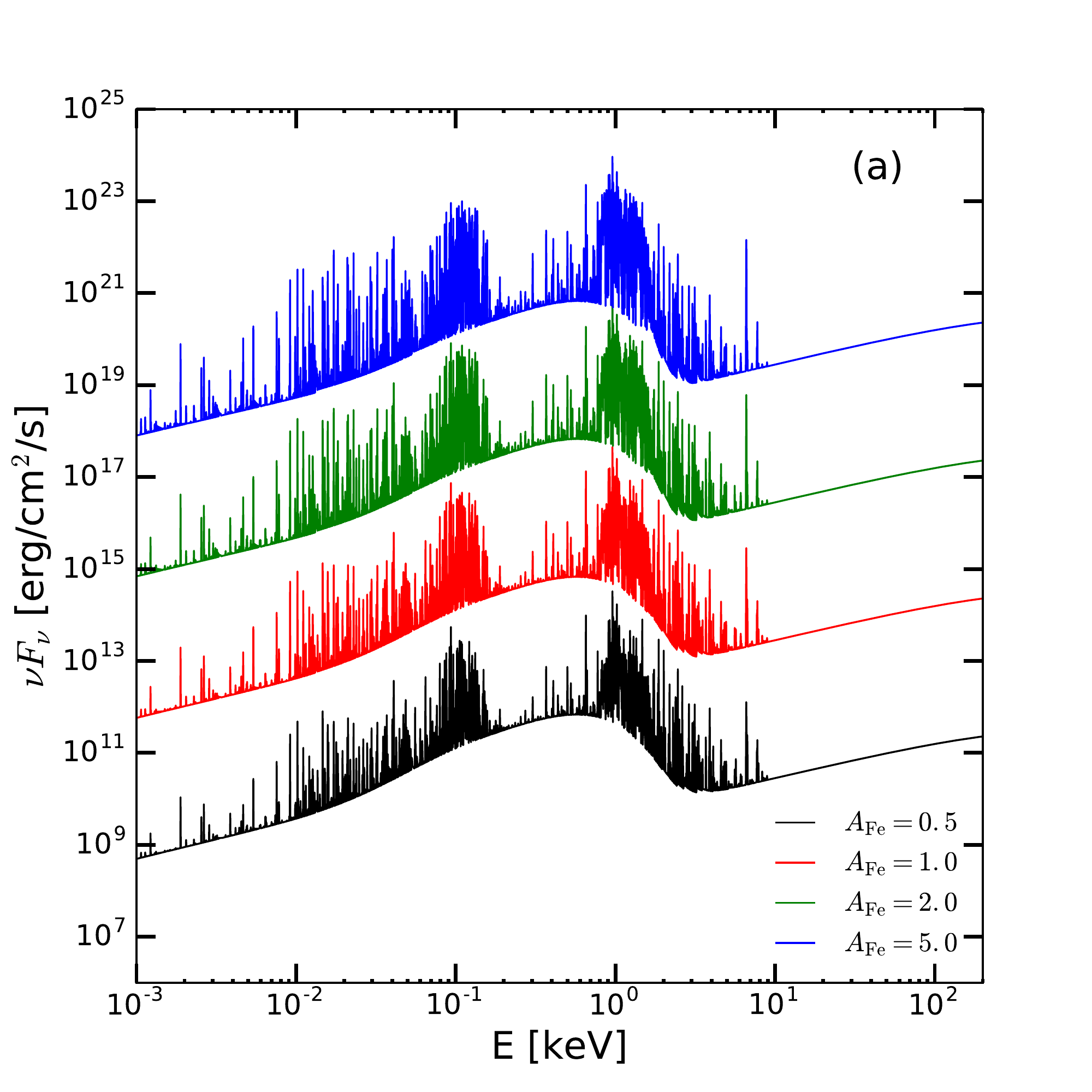}
    \hspace{-0.5cm}
    \includegraphics[height=8.0truecm]{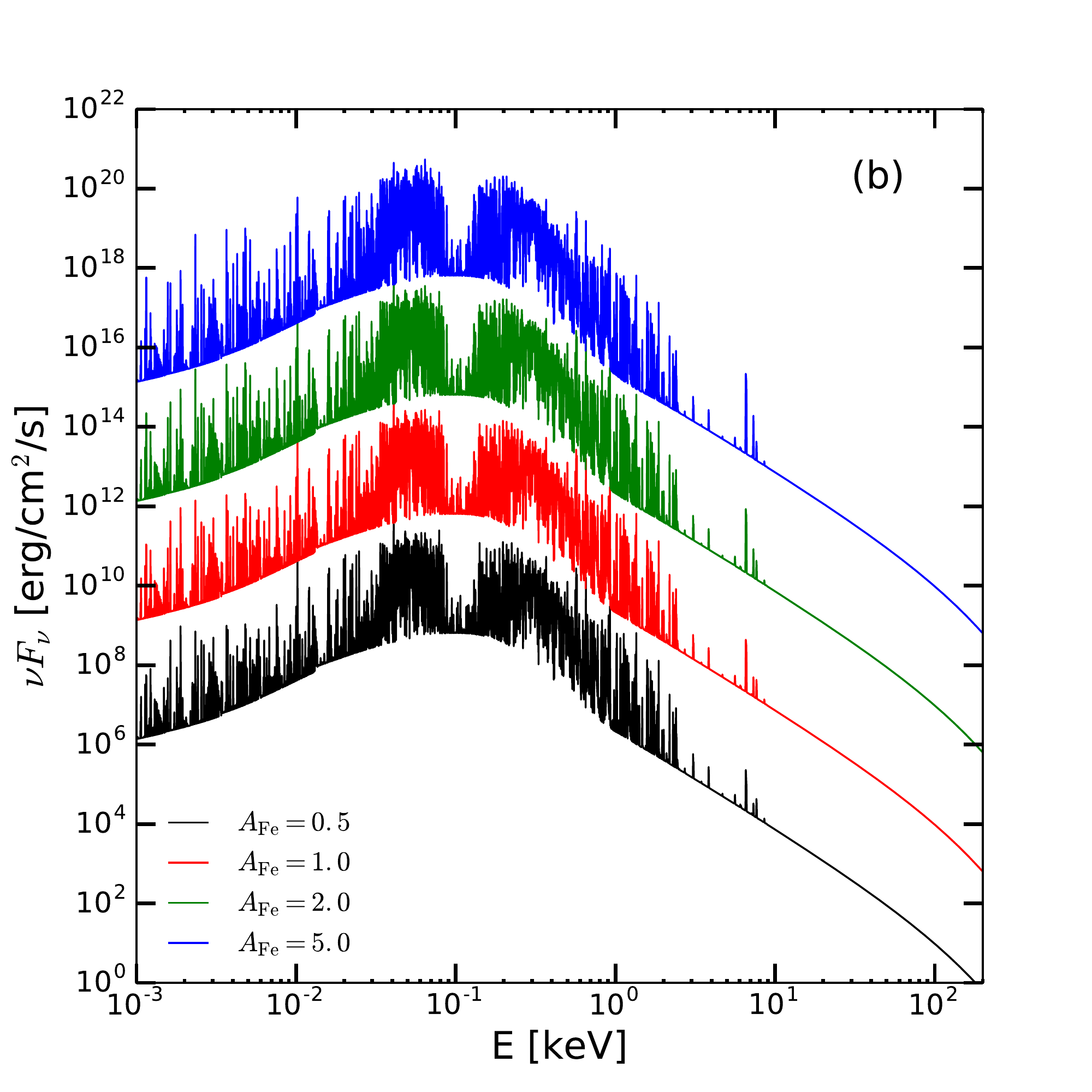}}
    \caption{Reflected spectra for different values of $A_{\rm Fe}$, with $A_{\rm Fe}$ = 0.5, 1.0, 2.0 and 5.0 when $\log \xi$=3. The TCAF model parameters which are fixed for both cases are, $M_{\rm BH}$=7.0~M$_\odot$, $X_{\rm s}$ = 30 and R = 2.5. The accretion rates for the two cases of spectral states are: (a) $\dot m_{\rm d}$ =0.001, $\dot m_{\rm h}$ = 1.0, and (b) $\dot m_{\rm d}$ =0.02, $\dot m_{\rm h}$ = 0.001. Successive spectra have been offset by factors of 1000 for clearer visibility.
}
\label{fig:reflection_AFe_mdmh}
\end{figure*}

The amount of a particular element present in the gas changes the continuum opacity, which in turn affects the photoionisation heating rate. Therefore the abundances for each element are taken into account in a photoionisation effect can greatly affect the ionisation balance, the observable spectral line features in the reprocessed radiation. At the same time, the abundance of a particular element influences the strength of the emission lines. Considering the relevant effect of Fe line emission in the analysis of the X-ray spectra for accreting binary candidates, we have carried out generating emission spectra in which the Fe abundance is varied between sub-Solar, Solar, and super-Solar values. All other elements considered in these calculations are set to their Solar values. The effects of varying abundance of iron are shown in \autoref{fig:reflection_AFe_mdmh}(a-b). The illuminating radiation has $\log \xi$ = 3, and the other model parameters are the same as in \autoref{fig:reflection_xiAfe} for two different spectral states, one is a very hard state when ARR=0.001 (left panel) and the other one is a softer state when ARR=20 (right panel), are shown in \autoref{fig:reflection_AFe_mdmh}a and b respectively. In each panel, the plotted spectra have been re-scaled for visual clarity. The scaling factors are, from bottom to top, 1, $10^3$, $10^6$, and $10^9$. The effect of $A_{\rm Fe}$ substantially changes the Fe emission features, evident in the emission spectra. For higher $A_{\rm Fe}$, increases the strength of overall emission lines including Fe emission. A significant difference at the low energy emission can be seen, where for higher ARR more lines are visible with higher intensity compared to low ARR. The reason behind this is that harder radiation (low ARR) fully ionised the disc therefore less lines are generated.

\subsection{Temperature profile and the effect of varying disc column density $N_{\rm H}$}
\begin{figure} 
    \centering{
    \includegraphics[height=6.0truecm]{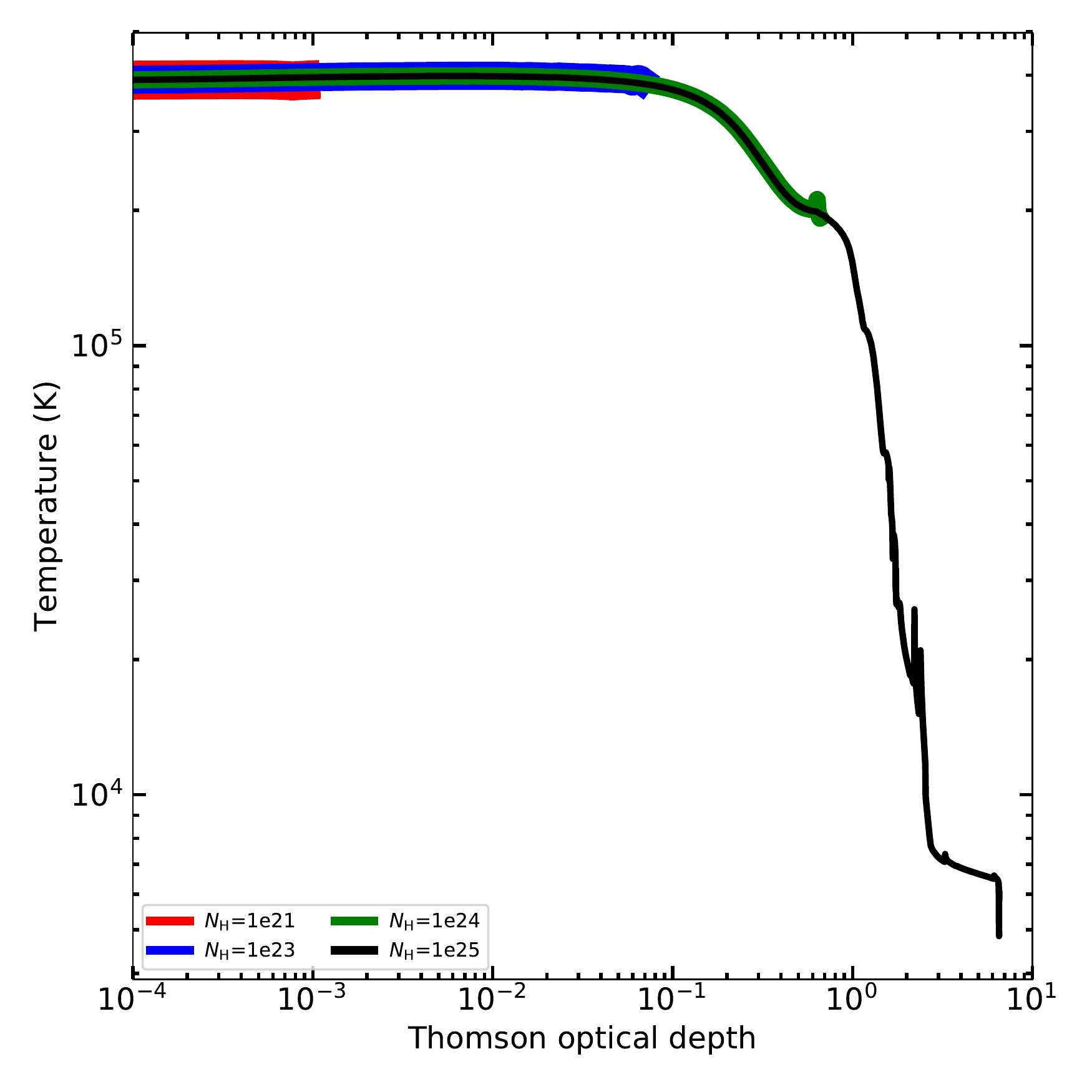}}
    \caption{Temperature as a function of the optical depth for models with variation in column densities. The structure presented here is for the incident SED generated using model parameters $\dot{m_{d}}=1.2$, $\dot{m_{h}}=1.0$, $X_{\rm s}$=30, and $R$=2.5. $\log \xi$ = 3.0 and Solar values for Fe abundances are used in {\sc cloudy}.}
    \label{fig:od}
\end{figure}

\begin{figure} 
    \centering{
    \includegraphics[height=6.0truecm]{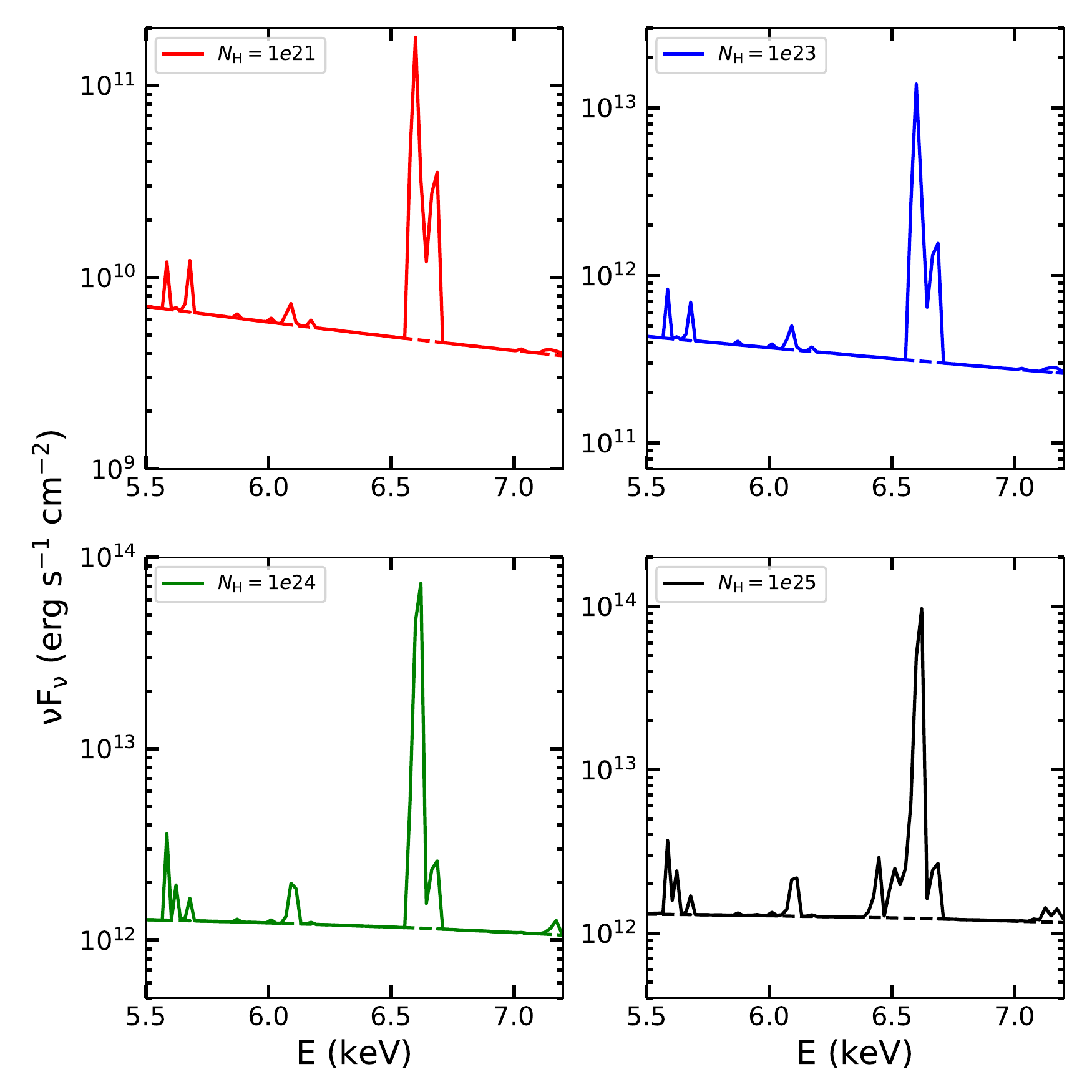}}
    \caption{Fe-line spectra for various values of $N_{\rm H}$ and for the same model parameters as in Fig.~\ref{fig:od}}
    \label{fig:Fe_column}
\end{figure}

We studied the dependence of our results on the thickness of the cloud by varying the column density of the cloud, $N_{\rm H}$. \autoref{fig:od} shows that increase in optical depth decreases the temperature of the illuminated medium for the disc parameter $\dot{m_{d}}=1.2$, $\dot{m_{h}}=1.0$, $X_{\rm s}$=30, and $R$=2.5.
This behaviour is due to the increased reprocessing of the X-ray photons by the larger surface of the gas cloud. The column density and optical depth considered here can potentially affect the temperature structure of the gas cloud by varying heating and cooling efficiency, thus the strength of the emission. At the surface of the gas layer where illuminating radiation hits is hot and Compton heating and cooling dominates. The temperature at the surface remains at Compton temperature. In this region radiation field thermalises, thus the temperature of the gas remains constant $\sim 4\times10^{5}$~K. As the optical depth increases, cold region exists and the temperature decreases, after a certain optical depth ($\sim 0.05$), the temperature falls sharply by orders in magnitude. If the column density is typically low, the cooling rate is also low, thus the illuminating radiation thermalises the layers of the cloud, thus the temperature remains constant to a fixed value does not fall as we see in high column density case. The blue, orange, green, and red curves show temperature profile for the values of $N_{\rm H}$ = $10^{21}$, $10^{23}$, $10^{24}$, and $10^{25}$ respectively. However, the scenarios may change if a different set of input SEDs and ionisation parameter are considered.
\autoref{fig:Fe_column} shows the effect of $N_{\rm H}$ on reflection spectra of Fe line between 5.5-7.2~keV, where line intensity increases with increasing column density. It also shows that when intensity increases, line width increases as evident from \autoref{fig:ew}. The change in line shape and the line to continuum contrast on increasing the column density are related to the behaviour of the curve of growth (COG) (i.e., the plot of line $W_E$ against the ionic column density). The curve of growth generally has three parts : linear part (where $W_E$ increases linearly with ionic column density), exponential part ($W_E$ increases exponentially), and saturated part (lines are saturated). Such a behaviour can be seen in \citet{Adhikari2019b}. For the lowest column density, the emission lines are produced corresponding to the ionic column density at the linear part of the COG. Once the column density increases, the lines are produced from an exponential part of the COG, hence, having higher $W_E$ and broader lines. Due to the higher ionic column density, the collisional broadening becomes more significant and the line shape becomes complex. If we keep increasing the column density further, it should remain unchanged as we approach the saturated part of COG.  
The Fe~XXVI (6.96 keV) line will show up for the cases where $\dot m_{\rm d}$ is low i.e. more hard photons can contribute and also the ionization rate is high. Therefore \autoref{fig:Feline_xiAfe}a shows the Fe~XXVI line up, whereas the panel b does not show up as the $\dot m_{\rm d}$ is high and $\dot m_{\rm h}$ is low (typical soft-state). Since \autoref{fig:Fe_column} is shown for the $\dot m_{\rm d} = 1.2$ (which lacks hard X-ray photons), hence the Fe~XXVI line is not present there. Furthermore, the 5.6~keV features are not Fe-line features (these are Cr and Ti line features not relevant at present). In addition the 6.1 (6.07-6.11) keV features for Mn, Ti, Cr, are not relevant at present as well. We put them for the sake of completeness and in the future the high-resolution observations may detect these lines.

\section{Shape of Fe Line} 
\label{sec:Fe_Shape}
\begin{figure} 
    \centering{
    \includegraphics[height=8.0truecm]{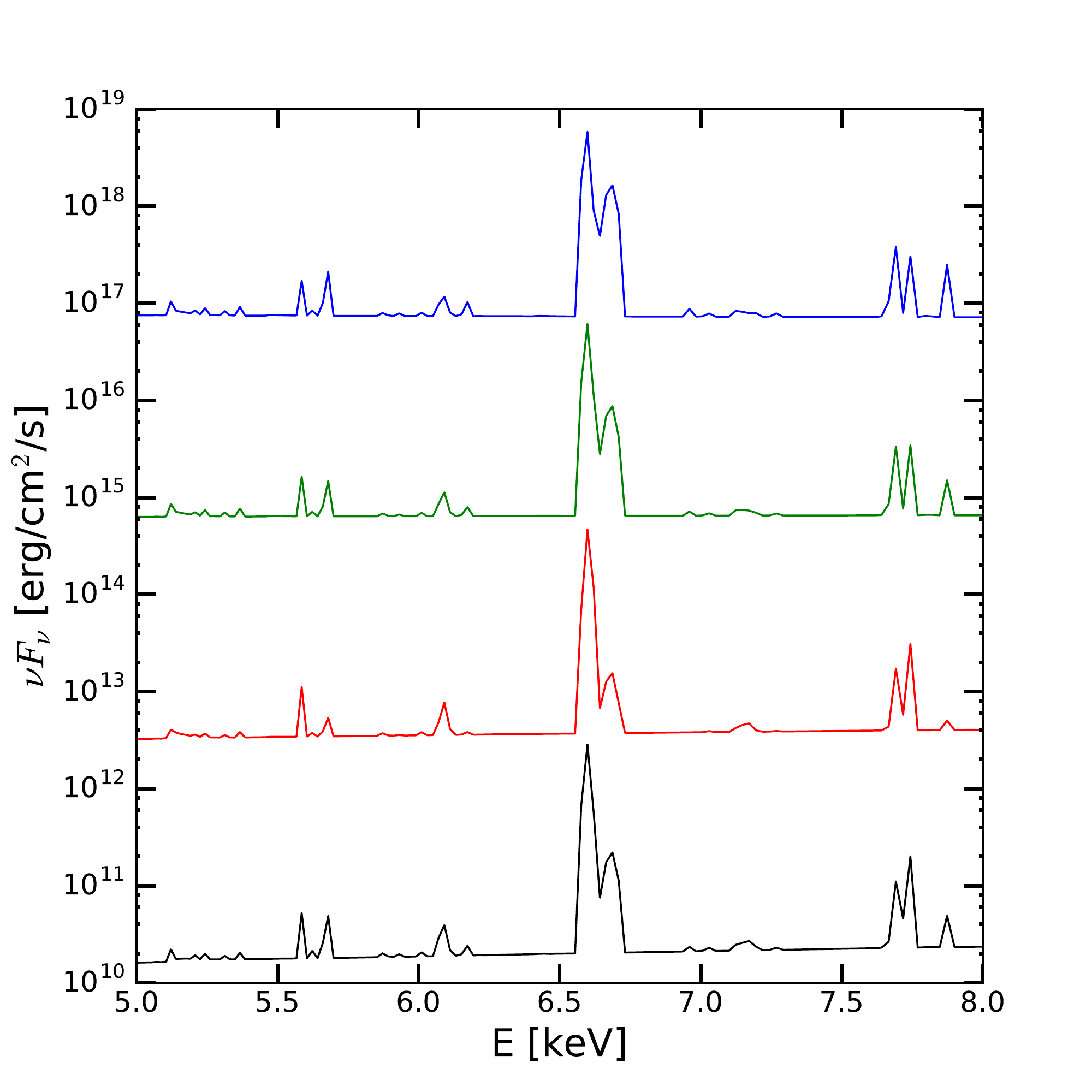}}
    \caption{Same as \autoref{fig:reflection_md}, but for the reflected spectra of the Fe lines. We use offset 100 for all Fe line spectra for visual clarity.}
    \label{fig:Feline_md}
\end{figure}

In \autoref{fig:Feline_md}, we show Fe line emission for varying $\dot m_{\rm d}$, same as in \autoref{fig:reflection_md}. Increasing disc mass accretion rate, the observed 6.6~keV and 6.7~keV double lines intensity increases. In all four cases, Fe line $\sim 6.6$~keV splits into two and many other lines from low Z species are produced between 5-8~keV energy band, some of them are mentioned in \autoref{fig:Fe_column}. In this case, ARR value increases refers increase in disc rate i.e. no of soft photon increases keeping the hot flow component (halo rate) fixed. Therefore spectra become softer. As the ARR value increases the second peak intensity increases and the first peak decreases. For high ARR, the outer layer of the gas cloud falls quickly from very hot phase to cold phase at negligibly small Thomson optical depth, therefore it can ionize more layer in the cloud to emit lines. 

\begin{figure} 
    \centering{
    \includegraphics[height=8.0truecm]{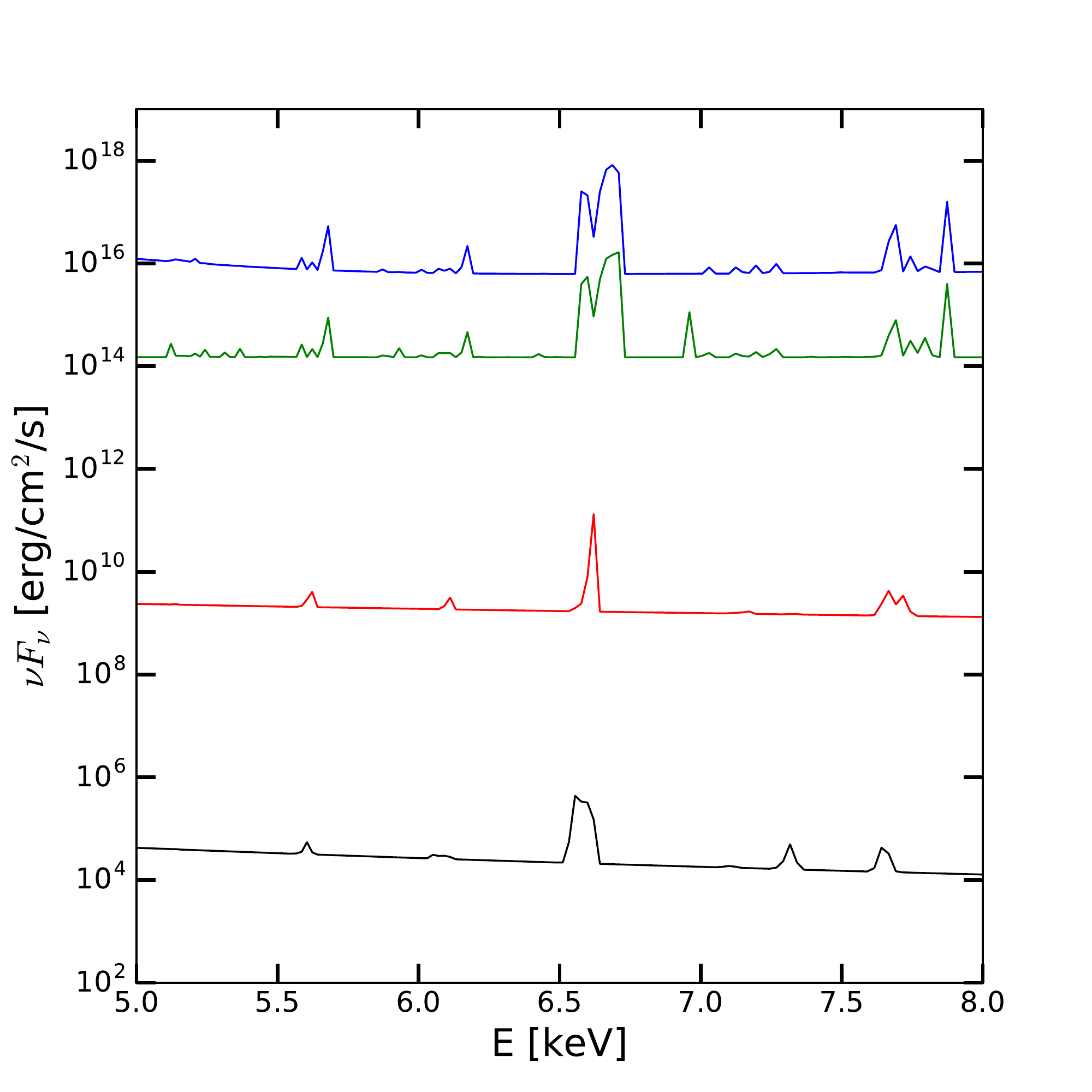}}
    \caption{Same as \autoref{fig:reflection_mh}, but for the reflected spectra of Fe lines.}
    \label{fig:Feline-mh}
\end{figure}

However, \autoref{fig:Feline-mh} shows a different line properties when halo rate increases from sub-Eddington to Eddington rate, Fe line $\sim 6.5$~KeV starts splitting and the component at $\sim 6.7$~keV has higher intensity than the lower energy peak. It is also noticeable that many complex lines are disappeared at low halo rate, and they appeared at high halo rate, this implies that when spectral states move from soft to hard, the intensity of illuminating spectrum at high energy increases, thus the ionisation rate, producing more lines. Here, increase in hot flow rate refers decrease in ARR which generate more hard photons to illuminate the disc surface. Therefore, at high ARR ( black and red lines) the illuminating radiation has not enough high energy photons to trigger thermal instability as the temperature of the gas cloud decrease, therefore less no of Fe species contributed and showed a single sharp peak. As the ARR goes down ($<$1, green and blue lines), the temperature of the gas cloud increases, the Fe line broadens. 

\begin{figure*} 
    \centering{
    \includegraphics[height=8.0truecm]{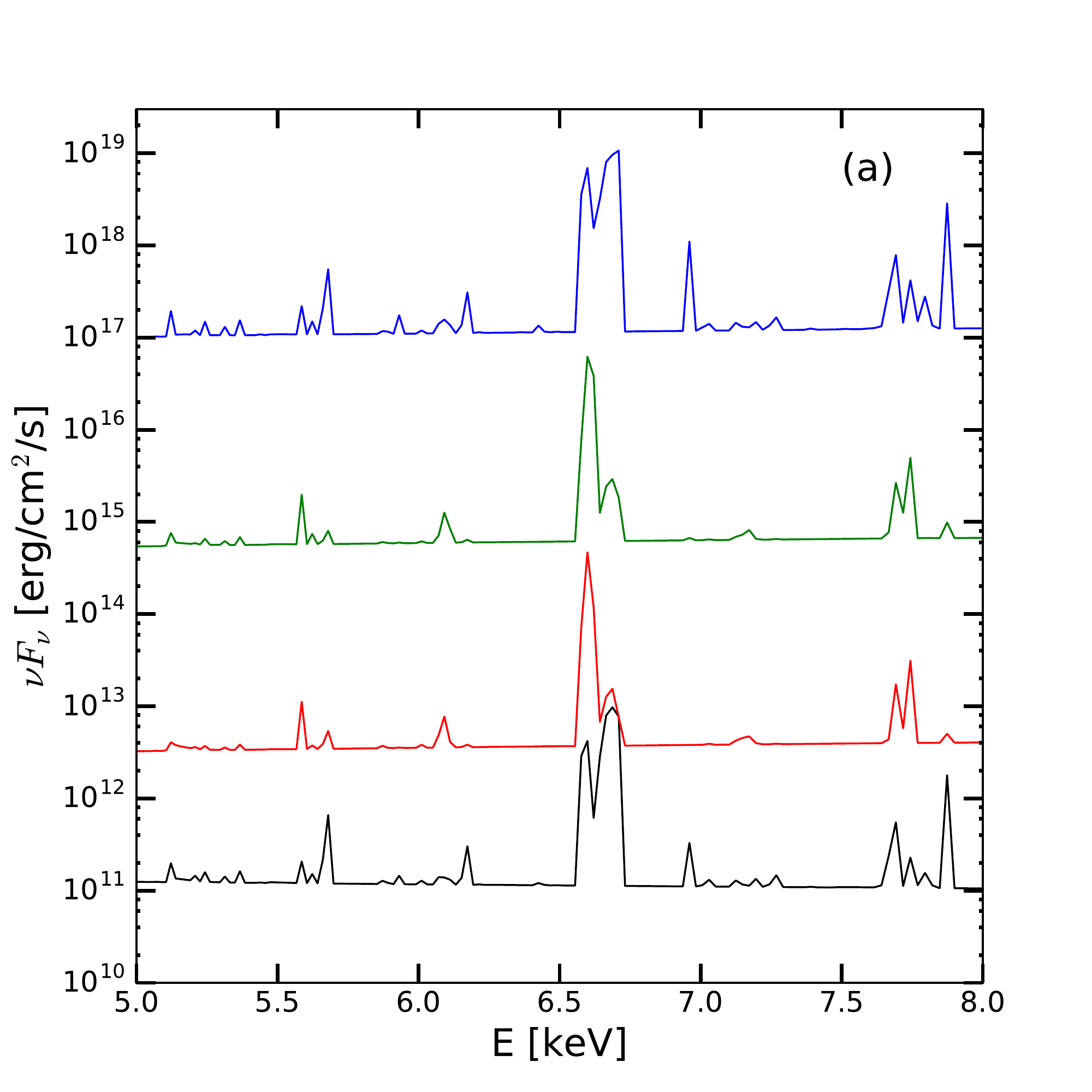}
    \hspace{-0.5cm}
    \includegraphics[height=8.0truecm]{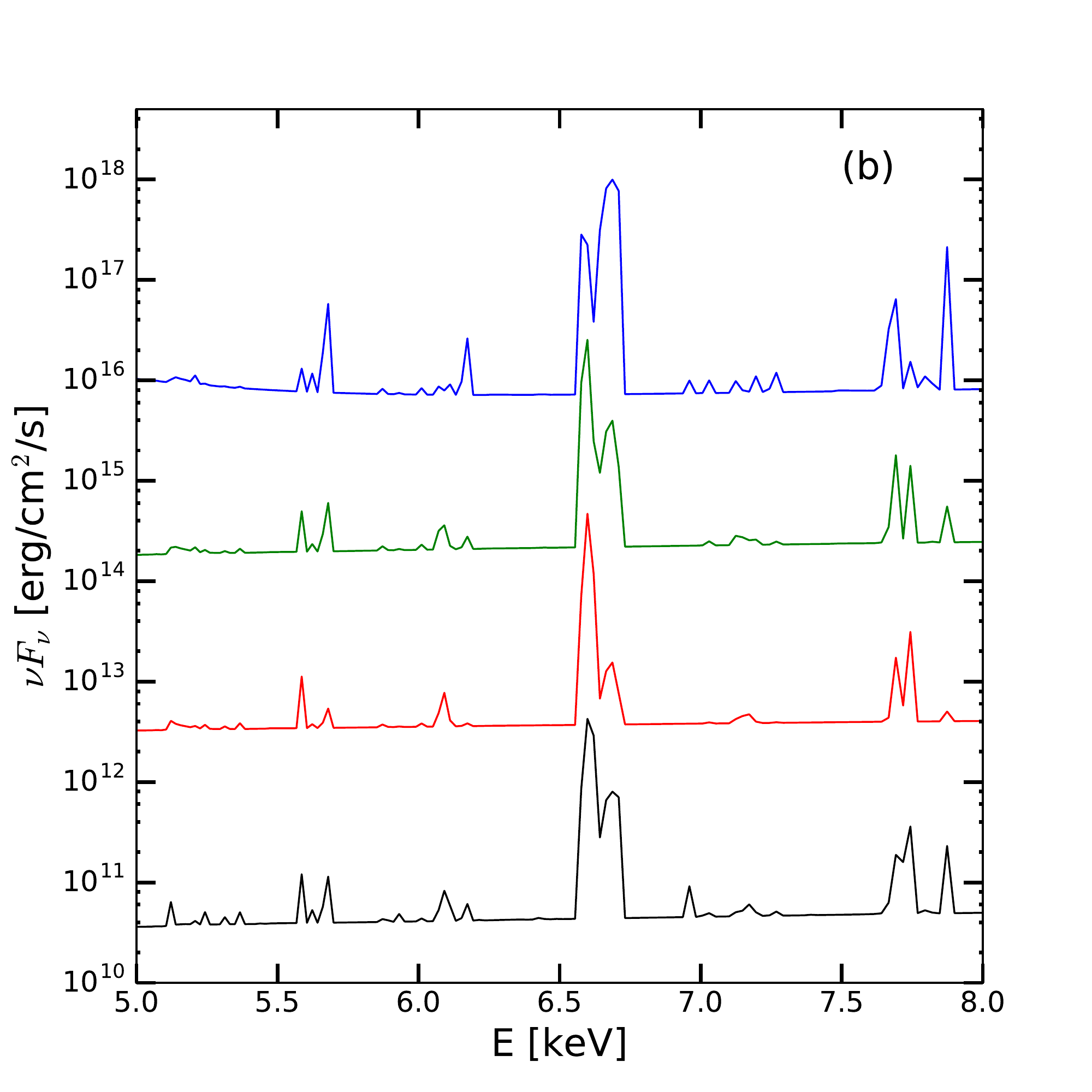}}
    \caption{Same as \autoref{fig:reflection_xs}, but for the reflected spectra of Fe lines.}
    \label{fig:Feline_xsR}
\end{figure*}

\autoref{fig:Feline_xsR}a-b show the effect of varying $X_{\rm s}$ and R on the reflected Fe line spectra. In both cases, Fe lines showed a double peak $\sim$6.5~keV. \autoref{fig:Feline_xsR}a shows that increasing $X_{\rm s}$, the low energy peak intensity of Fe lines increases, however, in low and high $X_{\rm s}$, lines showed opposite nature compared to the intermediate corona size. Also, an extra emission line $\sim$7~keV is observed in low and high $X_{\rm s}$ with the increase in intensity for high $X_{\rm s}$, these can be understood from the hardening of the illuminating spectra, and more number of hard photons are contributing in ionizing the disc. \autoref{fig:Feline_xsR}b, showed an opposite behavior in the double peak Fe line, where the intensity of the low energy peak $\sim$6.6~keV increases with R, and the high energy peak intensity did not change much, which showed flipping behavior for $R$=4, the strong shock condition. It should be noted that the double peaks which are observed are not due to any gravitational broadening effects nor due to photon bending effects, rather two different transitions of Fe lines at different energies.
The low energy iron lines are not shown up, which can be due to completely stripped off of the electrons from the atomic levels.


\begin{figure*} 
    \centering{
    \includegraphics[height=8.0truecm]{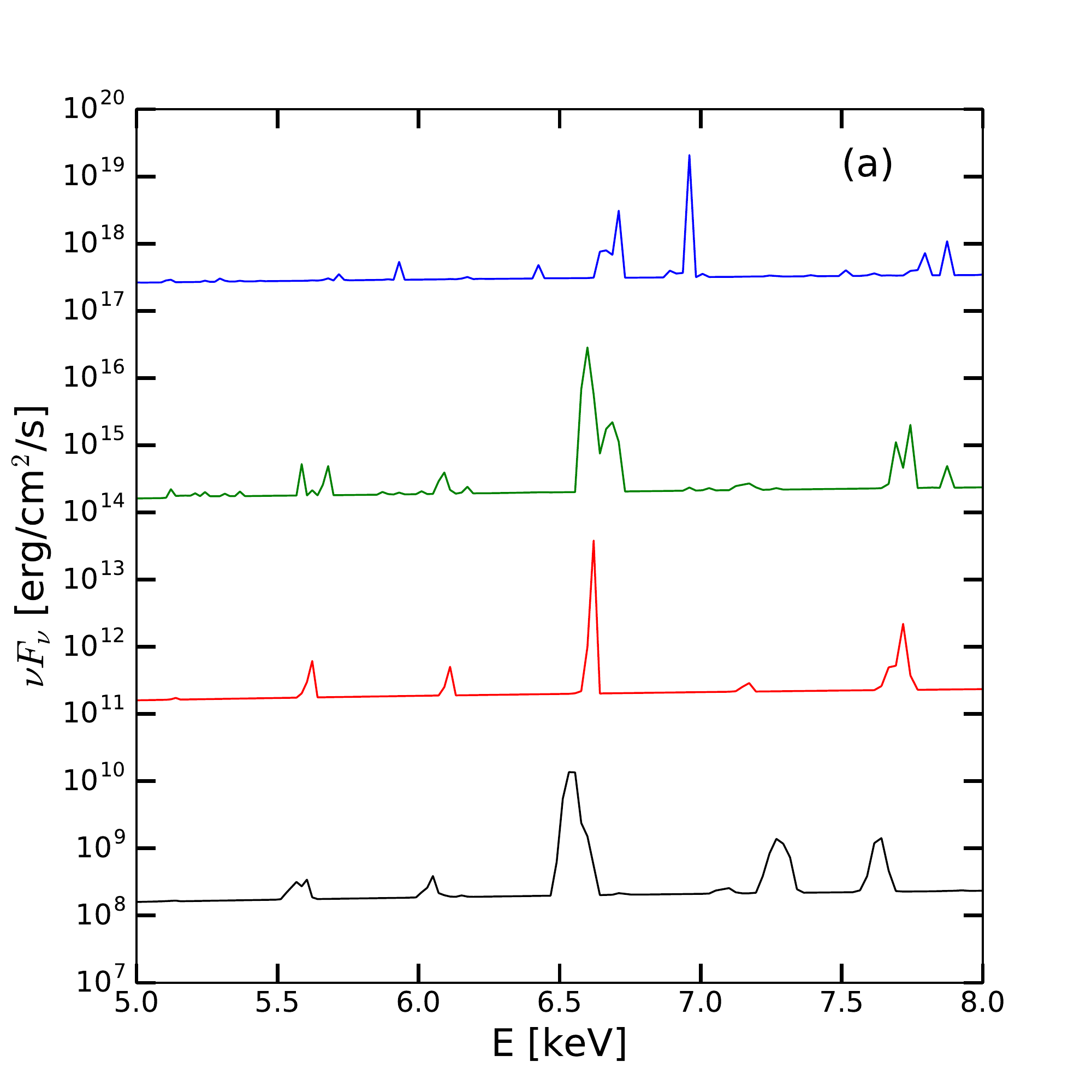}
    \hspace{-0.5cm}
    \includegraphics[height=8.0truecm]{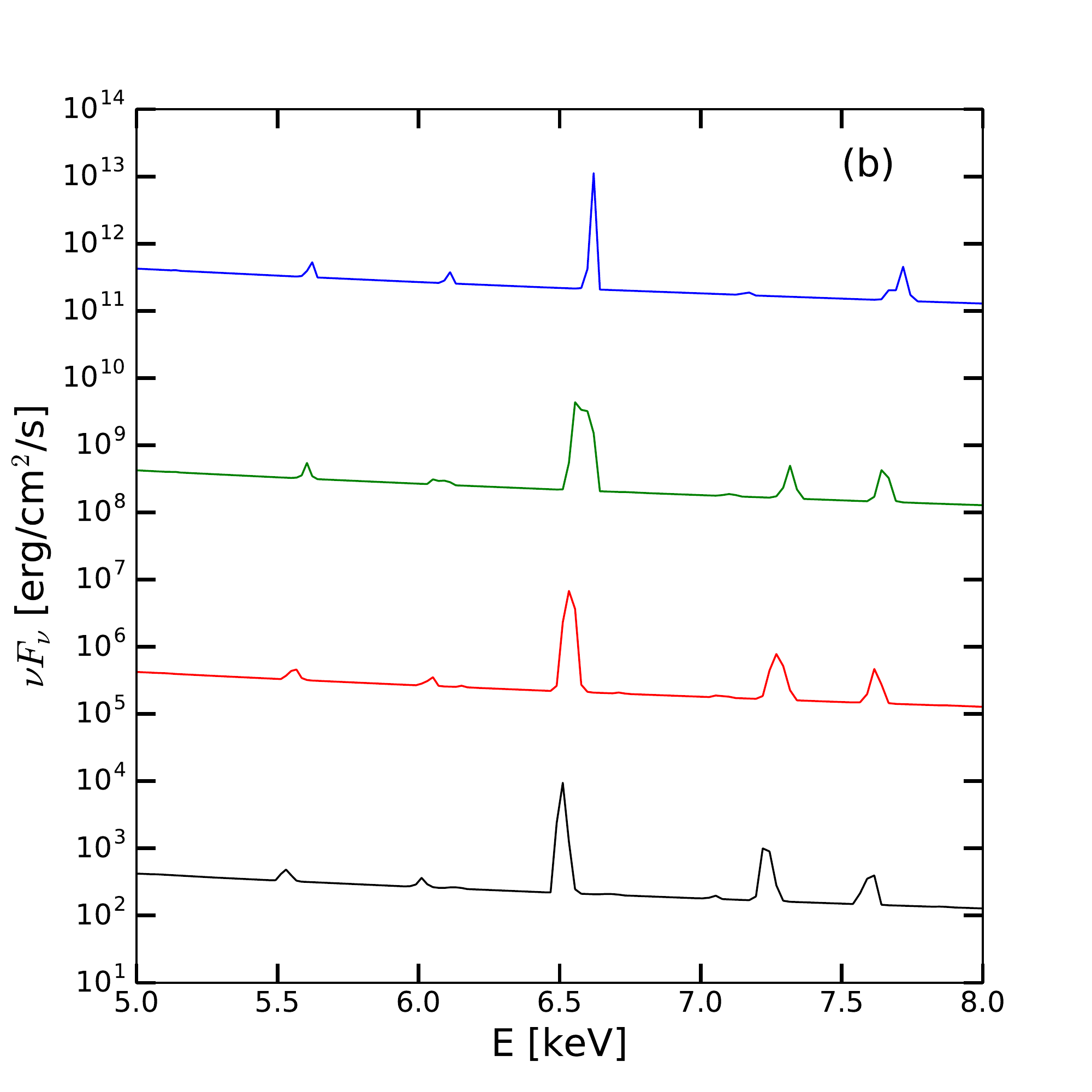}}
    \caption{Same as \autoref{fig:reflection_xiAfe}. Reflected spectra for Fe lines.}
    \label{fig:Feline_xiAfe}
\end{figure*}

\autoref{fig:Feline_xiAfe} a-b show the effect of varying $\log \xi$ on the reflected Fe line spectra for the same sets of model parameters in \autoref{fig:reflection_xiAfe} for HS and SS. At the pure HS with low $\log \xi$, a single and broad Fe line is observed at $\sim 6.5$~keV. It becomes narrow when $\log\xi$ increases and the line splits when $\log \xi$ is 3, at very high ionisation value the line intensity goes down and an another sharp-peaked line at 6.96~keV with high intensity is observed. This evidences that the gas becomes so ionized that H like iron (Fe XXVI) is the dominant species. However, for the same $\log \xi$, in the SS \autoref{fig:Feline_xiAfe}b, only one line is observed at 6.5~keV and it is narrower in low and high $\log \xi$. The appearance of the single peak line in the SS follows the similar discussion of temperature structure change of the gas cloud with the hardness of the illuminating radiation. As the ARR is high ($\gg 1$), incident radiation has not enough hard photons to trigger the thermal instability, therefore not many Fe species contributed in line emission.

\begin{figure*} 
    \centering{
    \includegraphics[height=8.0truecm]{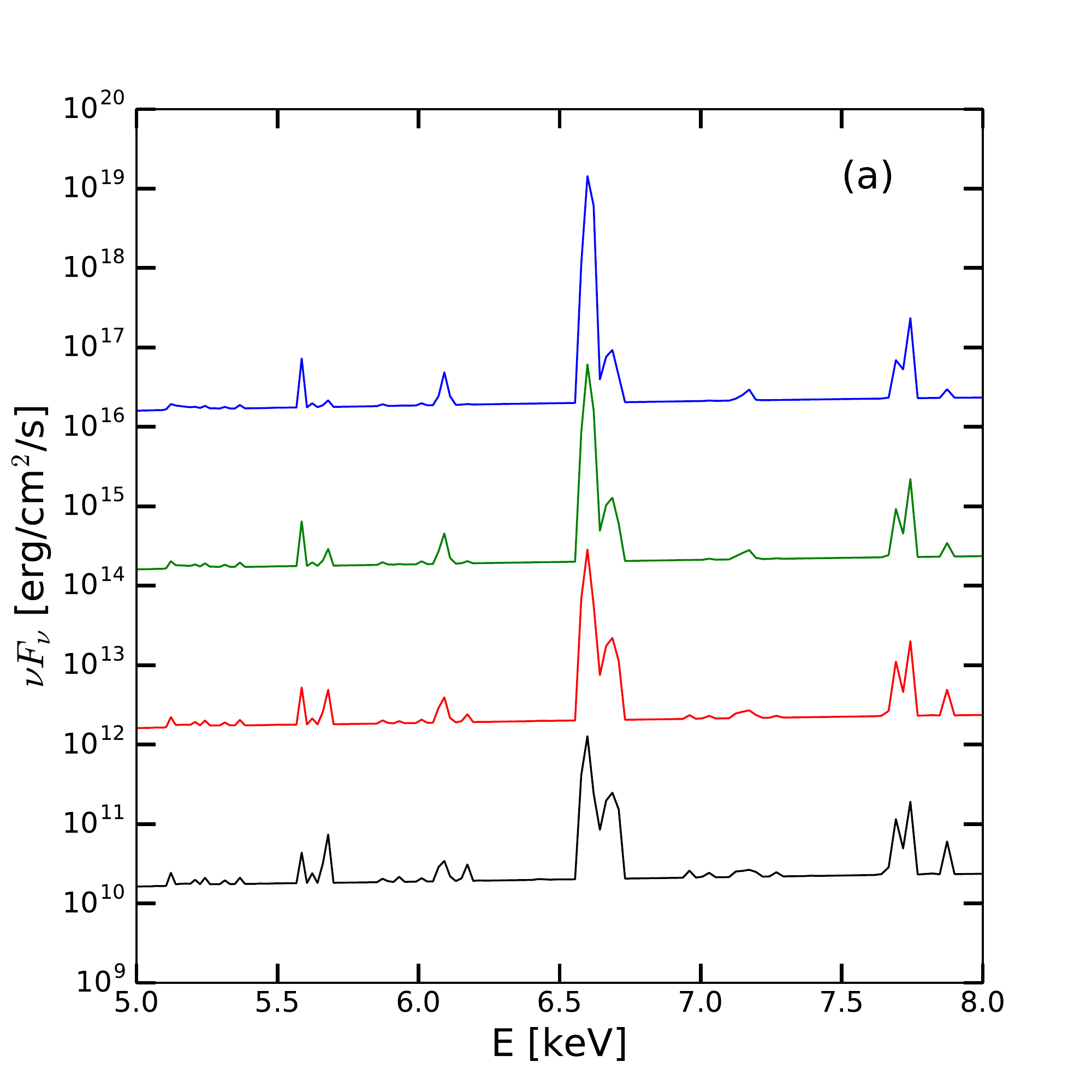}
    \hspace{-0.5cm}
    \includegraphics[height=8.0truecm]{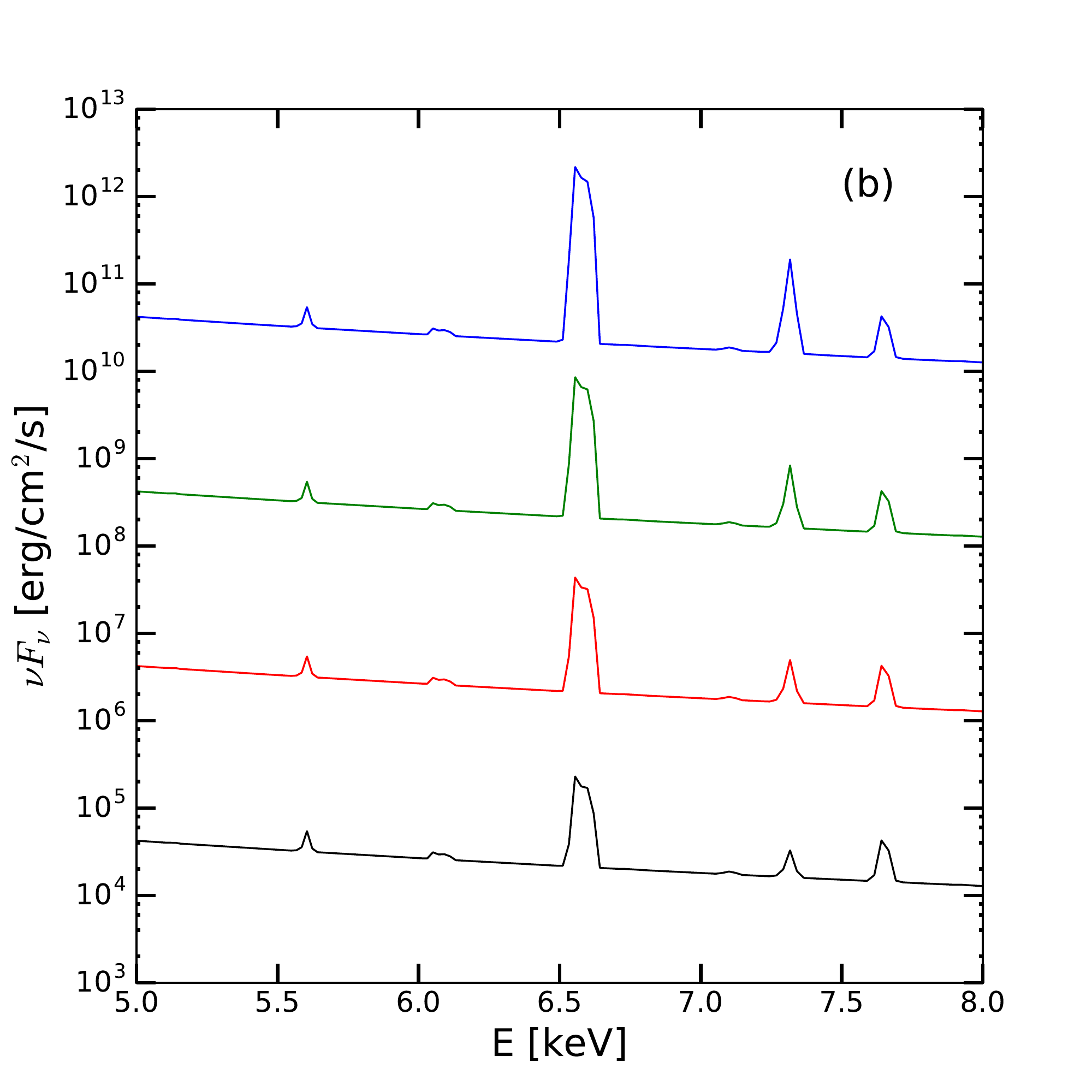}}
    \caption{Same as \autoref{fig:reflection_AFe_mdmh}. Reflected spectra for Fe lines.}
    \label{fig:Feline_mdmh}
\end{figure*}

\autoref{fig:Feline_mdmh}a-b show the effect of varying $A_{\rm Fe}$ (increases from black to blue line) on the reflected Fe line spectra for the same sets of model parameters in HS and SS as in \autoref{fig:reflection_AFe_mdmh}. \autoref{fig:Feline_mdmh}a shows the double peak nature of Fe lines with increasing intensity in low energy line, whereas high energy line intensity decreases, this is due to the hard radiation increases ionisation rate with increasing Fe abundances. Thus more complex lines are formed. However, \autoref{fig:Feline_mdmh}b shows a single line with increasing intensity for increasing $A_{\rm Fe}$, indeed shows a significant effect of Fe abundances. Also, the illuminating spectra have less effect on the temperature structure of the gas cloud in the SS (ARR $>$ 1) as less number of hard photons contributed to trigger the thermal instability. However, from observation, we see that the line broadens and breaks into a double peak during the outburst phase, and the intensity increases as the spectrum moves to the soft state \citep{Mondaletal2016} when the inner edge of the disc comes much closer to the BH. These two combined results infer that gravitational broadening might be more dominating over photoionisation in the soft state, when the Fe lines are originated at the inner edge of the disc. 

In the current modeling, we did not consider any gravitational effect as in \cite{Fabianetal1989} or \cite{Laor1991}. The lines are generated due to photo-ionization processes or due to atomic transitions in {\sc Cloudy}. It is also worth mentioning that {\sc Cloudy} does not have the gravitational broadening effect, but the broadening due to the thermal motion of the gas is incorporated. Furthermore, our location of the shock is not close to the innermost stable circular orbit, rather farther away, where the line broadening effect may not be effective enough. Moreover, general relativistic or photon bending effects are important to get those double-peaked lines, which is beyond the scope of the paper.

\section{Conclusions} \label{sec:conclusion}
In this paper, we have presented reprocessed spectra from an accretion disc, which is emitted as a reflected spectra. The incident illuminating spectra are generated using TCAF model. Each model is characterized by the disc and halo mass accretion rates, location of shock/size of the corona and the shock compression ratio of the flow. Next, the reprocessing of this incident radiation is simulated by using the photoionisation code {\sc cloudy}, where a disc is parametrized by: the ionisation parameter at the surface, the hydrogen number density, the column density and the Fe abundance with respect to its Solar value. The range of parameters we use here, covers the observed spectral states for accreting black hole binaries. The parameter ranges are following: $0.001 \le \dot m_{\rm d} \le2.0$, $0.001 \le \dot m_h \le1.5$, $10\le X_{s} \le 100$, $1.5\le R \le4.0$, $1\le \log \xi \le 4$, and $0.5\le A_{\rm Fe} \le 5$. The illuminating spectra consist disc blackbody and the Comptonization component of the spectra taking into account the reflection, covering the energy range from 0.001 to 1000~keV. 

In comparison to other line emission models in the literature, our illuminating spectra vary in a multidimensional parameter space including the dynamics of the flow, which is indeed observed for outbursting black hole candidates. The key findings of the present model are as follows:
\begin{enumerate}
    
    \item Increasing $\dot m_{\rm d}$, the spectrum becomes softer, the intensity of the emission lines at the high energy part of the spectrum decreases, but more lines are observed at low energies. Varying $\dot m_{\rm d}$ splits Fe line into double peak $\sim 6.65$~keV (Fe XXV) with higher intensity at low energy line and many complex lines are formed (see \autoref{fig:Feline_md}). The Fe line intensity correlates with $\dot m_{\rm d}$.
    
    \item Changing $\dot m_{\rm h}$, covering from hard to soft followed by intermediate, shows that in the soft state more lines are observed throughout the energy band, however, in the intermediate state (red line in \autoref{fig:reflection_mh}) emission lines are clearly visible throughout the energy band. In hard state, more/less lines are visible at high/low energies. In the pure soft state and intermediate state (black and red lines in \autoref{fig:Feline-mh}), Fe lines have a single peak $\sim 6.65$~keV with higher equivalent line width in the soft state and a sharp peak in the intermediate state. The Fe line splits into a double peak when it enters into hard state with opposite behavior in peak intensity. 
    
    \item Increasing $X_{\rm s}$, low energy peak intensity of Fe line increases, whereas the peak intensity of high energy decreases. But both peaks are comparable for $X_{\rm s}=100$ case (see \autoref{fig:Feline_xsR}a) and a sharp peaked H like iron (Fe XXVI) line at $6.96$~keV has been observed.
    
    \item Increasing R, low energy peak intensity is higher than high energy peak. However, when shock gets stronger, the peak intensity nature becomes opposite. This can be due to, at strong shock spectrum becomes harder thus efficiency of illuminating the disc by high energy photons is more, which emits lines with higher intensity (see \autoref{fig:Feline_xsR}b).
    
    \item In the soft spectral state, a single peaked Fe line is produced for different ionisation parameters, where the line width is less for the lowest and highest ionisation parameters compared to intermediate ionisation rate. In contrast, in the pure hard state, line splits at higher ionisation values, and at the highest ionisation value a high intensity, sharp peaked line at $6.96$~keV is observed. This implies that the gas become so ionized that H like iron (Fe XXVI) is the dominant species. This ion has a large fluorescent yield resulting in a strong line.
    
    \item A similar feature is observed when Fe abundance is varied. In soft spectral state, the intensity and width of the single peaked Fe line increase with Fe abundance, however, in pure hard state, the lines split into two and the intensity of the low energy peaks increase with increasing Fe abundance.
    
    \item The measurable quantity which relates the observation is the $W_{\rm E}$ of the Fe lines, which decrease with increasing disc accretion rate, and increases with the column density of the gas cloud. The equivalent width also decreases with ionisation parameter ($\xi$) up to a certain value ($\xi < 10^3$) and then again increases sharply with a peak $\sim 10^4$ and then falls sharply. 
    
    \item The temperature profile of the gas cloud above the disc changes by orders in magnitude with increasing optical depth, depending on the column density of the medium.
\end{enumerate}
In the present paper, we did not fit the simulated Fe lines directly with the observed lines. In the future we aim to directly fit the observed data using the current model as a table model or as a local model. The current model also does not include the inclination effect in the theoretical model generation, which will be published elsewhere.

\section*{Acknowledgements}
We thank the anonymous referee for making critical comments that improved the quality of the manuscript. We thank Agata ~R\'o\.za\'nska for insightful comments that improved the presentation of the manuscript. SM acknowledges Ramanujan Fellowship (file \#RJF/2020/000113) by SERB-DST, Govt. of India and Kreitman Fellowship by Kreitman School of Advanced Graduate Studies at the Ben-Gurion University of the Negev, Israel, during which the work was started. T. P. A gratefully  acknowledges the Inter-University Center for Astronomy and Astrophysics (IUCAA), Pune, India and the Nicolaus Copernicus Astronomical Center of the Polish Academy of Sciences, Poland for providing the access to the Computational Cluster, where the numerical simulations used in this paper are performed. C.B.S. is supported by the National Natural Science Foundation of China under grant No. 12073021.

\section*{Data Availability}
Data information may not be applicable for this article. No new data has been analyzed as it is mostly a theory-based article. The numerically created data that support the findings of this study are available from the corresponding author, upon reasonable request.



\bibliographystyle{mnras}
\bibliography{disc} %
\appendix

\bsp	
\label{lastpage}
\end{document}